\documentclass[prd,aps,floats,preprint,preprintnumbers,amssymb,longbibliography,nofootinbib]{revtex4-1}
\usepackage[utf8]{inputenc}
\usepackage{amsmath,amssymb}
\usepackage{graphicx}
\usepackage{hyperref}
\usepackage{physics}
\usepackage{bm}
\usepackage{caption}
\usepackage{float}

\usepackage{etoolbox}   
\usepackage{ragged2e}   

\usepackage{xcolor}
\definecolor{linkblack}{RGB}{20,20,20}
\definecolor{citeblue}{RGB}{0,70,140}
\definecolor{urlblue}{RGB}{0,90,120}
\usepackage{hyperref}
\hypersetup{
  colorlinks=true,
  linkcolor=linkblack,
  citecolor=citeblue,
  urlcolor=urlblue
}

\makeatletter
\patchcmd{\@makecaption}{\centering}{\justifying}{}{}
\patchcmd{\@makecaption}{\center}{\justifying}{}{} 
\makeatother

\newcommand{\eq}[1]{Eq.~\ref{#1}}

\begin{document}

\title{The fate of Schwarzschild–de Sitter black holes: \\ nonequilibrium evaporation}

\author{Damien A. Easson}\email{easson@asu.edu}
\affiliation{Department of Physics, Arizona State University, Tempe, Arizona 85287, USA}
\affiliation{Beyond Center for Fundamental Concepts in Science, Arizona State University, Tempe, Arizona 85287, USA}

\begin{abstract}
We present a fully analytic treatment of Schwarzschild--de~Sitter (SdS) black-hole evaporation in two-dimensional dilaton gravity with anomaly-induced backreaction. Starting from the spherical reduction of four-dimensional Einstein gravity with a cosmological constant, we construct an exactly solvable 2D model that captures the full causal and thermodynamic structure of the SdS static patch, including both black-hole and cosmological horizons. Incorporating the trace anomaly of $N$ conformal matter fields via the Polyakov action, we determine the evolution of the black-hole mass and geometry in the Unruh--de~Sitter state, track the steady nonequilibrium Hawking flux, and compute local thermodynamic observables for static observers. The conserved Killing energy flux drives an irreversible heat current from the black hole to the cosmological horizon whenever their surface gravities differ, ensuring monotonic entropy growth and satisfaction of the generalized second law. We prove that $\kappa_b>\kappa_c$ throughout the physical static patch, so the only zero-flux configuration is the Nariai limit where the horizons coincide. Extending the framework to the quantum-information regime, we construct a thermo-controlled estimate of the Page curve and show how quantum extremal surfaces and entanglement islands emerge naturally within the anomaly-induced steady state. These results constitute a fully analytic, backreacted solution for SdS evaporation that unifies semiclassical thermodynamics and information flow in a cosmological setting, thereby elucidating the ultimate fate of evaporating black holes in de~Sitter space.
\end{abstract}

\maketitle
\tableofcontents

\section{Introduction}

The discovery by Bekenstein and Hawking that black holes radiate thermally and
carry entropy revolutionized our understanding of gravity and thermodynamics,
revealing that classical event horizons are inherently quantum
objects~\cite{Bekenstein:1973ur,Bekenstein:1974ax,Bardeen:1973gs,
Hawking:1974rv,Hawking:1975vcx}.
In an accelerating de~Sitter universe, black holes coexist with a cosmological
horizon, each characterized by its own surface gravity and associated
temperature.
Unlike asymptotically flat black holes, this dual-horizon structure places the
system in intrinsic nonequilibrium: both horizons radiate, but generally at
different temperatures.
Despite significant progress, a complete understanding of black-hole evaporation
in de~Sitter space and of how the second law of thermodynamics is realized in such
spacetimes remains a fundamental open problem.

A powerful framework for addressing semiclassical backreaction is provided by
the two-dimensional trace-anomaly method pioneered by
Polyakov~\cite{Polyakov:1981rd} and developed in many subsequent
works~\cite{Christensen:1977jc,Davies:1976ei,Callan:1992rs,Russo:1992ax,
Fabbri:2005mw}.
In this approach, the backreaction of $N$ conformal matter fields is encoded in an
anomaly-induced effective action that captures quantum stress--energy effects in
a geometrically transparent form, allowing controlled calculations of Hawking
flux and backreaction in both asymptotically flat and cosmological settings
(see, e.g.,~\cite{Callan:1992rs,Russo:1992ht,Bousso:1997cg,Balbinot:1998yh}).
Yet an analytic, self-consistent treatment of Schwarzschild--de~Sitter (SdS)
evaporation that simultaneously tracks both horizons, real-time mass evolution,
and the thermodynamic response of local observers has not yet been achieved.

Two-dimensional dilaton gravity has long served as a laboratory for exploring
black-hole thermodynamics, semiclassical backreaction, and departures from
classical singular behavior, including early studies of Hawking radiation,
nonsingular geometries, and regular Schwarzschild and
Schwarzschild--de~Sitter solutions~\cite{Poisson:1990eh,Banks:1992xs,
Trodden:1993dm,Bogojevic:1998ma,Easson:2001qf,Easson:2002tg,Easson:2017pfe,
Frolov:2021kcv,Davies:2024ysj}.
While these works provide important conceptual and technical insights into
regularity, backreaction, and effective descriptions of black-hole interiors,
they do not incorporate the full SdS static patch or yield a closed-form,
globally backreacted, two-horizon evaporation solution.

In parallel, four-dimensional studies of Schwarzschild--de~Sitter and
Kerr--de~Sitter evaporation have analyzed Hawking emission using greybody
factors and mode-by-mode fluxes for scalar, electromagnetic, and gravitational
perturbations~\cite{Bousso:1997wi,Kanti:2002ge,Yoshida:2003zz,Gregory:2021ozs},
demonstrating consistency with the generalized second law.
Related semiclassical analyses have constructed perturbatively
self-consistent SdS solutions by solving backreaction equations for quantum
fields on a fixed background~\cite{Akhmedov:2024mkk}.
However, these approaches rely on numerical computation or perturbative control
and do not yield closed-form expressions for the evaporation fluxes, mass
evolution, or local thermodynamic observables.

Here we construct an analytic framework that addresses these limitations.
Starting from the spherical reduction of four-dimensional Einstein gravity with a
cosmological constant, we obtain an exactly solvable two-dimensional dilaton
gravity model that reproduces the causal and thermodynamic structure of the SdS
static patch.
Incorporating the Polyakov effective action, we determine the anomaly-induced
energy flux, derive the evolution equation for the black-hole mass, and compute
local temperatures, energy densities, and fluxes measured by static observers.
In two dimensions, the anomaly-induced action captures the complete semiclassical
dynamics of the $s$-wave sector~\cite{Polyakov:1981rd,Russo:1992ax,
Alonso-Serrano:2025lbo}, enabling a self-consistent treatment of backreaction.

The resulting system realizes a nonequilibrium steady state with constant energy
flow from the hotter black hole to the cooler cosmological horizon.
The mass-loss law admits a single equilibrium point, the Nariai configuration,
where the horizons coincide and their surface gravities vanish.
We construct the $(M,\Lambda)$ phase diagram, identify its fixed points, and
verify that the generalized second law is satisfied throughout the evaporation
process.

Quantum extremal surfaces (QES) and entanglement islands in de~Sitter--like
settings have been analyzed previously in simplified two-dimensional models,
most prominently in the near-Nariai JT limit~\cite{Aalsma:2019rpt,
Aalsma:2021bit} and in related treatments of cosmological or near-extremal
throats~\cite{Hartman:2020khs,Shaghoulian:2022fop}, as well as in studies of
vacuum-dependent evaporation and island formation in near-extremal RNdS or
JT-like reductions~\cite{Bhattacharjee:2020nul}.
These works exploit the analytic tractability of near-horizon geometries, where
the dilaton takes a simple form and the generalized entropy can be extremized
explicitly.

Our framework differs in several essential respects.
We employ the full spherical reduction of four-dimensional SdS and incorporate
anomaly-induced backreaction throughout the entire static patch, rather than
restricting to a near-horizon JT limit.
In our Eddington--Finkelstein formulation, the Unruh--de~Sitter (UdS) state is
implemented through constant state functions $t_\pm$ (equivalently the conserved
flux $\mathcal J$), so the backreacted solution satisfies the UdS regularity
conditions without requiring an additional state-defining coordinate
transformation.
Consequently, our discussion of islands emphasizes the thermodynamic necessity of
the Page transition in a global nonequilibrium steady state, rather than a
microscopic extremization of $S_{\rm gen}$ in a near-horizon throat.
A detailed QES extremization in the fully backreacted SdS geometry is left for
future work.

The remainder of this paper is organized as follows.
Section~\ref{sec:reduction} reviews the spherical reduction to two-dimensional
dilaton gravity.
Section~\ref{sec:framework} develops the semiclassical anomaly framework and the
mass-loss law.
Section~\ref{sec:evap} analyzes evaporation dynamics,
Section~\ref{sec:exactsteady} the quasi-stationary geometry, and
Section~\ref{sec:equiconf} equilibrium and stability.
Sections~\ref{sec:odt} and~\ref{sec:GSL} discuss observer thermodynamics and the
generalized second law, followed by Section~\ref{sec:islands} on quantum extremal
surfaces, entanglement islands, and the Page curve.
Technical derivations appear in
Appendices~\ref{app:quadratures}--\ref{app:curvature}.

\section{From 4D Schwarzschild--de Sitter to 2D dilaton gravity}
\label{sec:reduction}
To fix notation and for completeness, we briefly summarize the spherical
reduction of four-dimensional Einstein--Hilbert gravity with a cosmological
constant to its two-dimensional dilaton formulation, following standard
treatments, and recover the Schwarzschild--de~Sitter (SdS) solution within the
reduced theory.\footnote{We closely follow the standard spherically reduced
gravity (SRG) formalism~\cite{Mann:1989gh,Klosch:1995fi,Grumiller:2002nm}.}

Starting from the 4D EH action with cosmological constant,
\begin{equation}
S_{4}=\frac{1}{16\pi G_{4}}\int d^{4}x\,\sqrt{-g^{(4)}}\left(R^{(4)} - 2\Lambda\right),
\end{equation}
we assume a spherically symmetric metric ansatz
\begin{equation}
ds_{(4)}^{2}=g_{ab}(x)\,dx^{a}dx^{b} + r(x)^2\,d\Omega_{2}^{2}, \qquad a,b\in\{0,1\},
\end{equation}
and define the dilaton as the area variable,
\begin{equation}
X(x):=r(x)^2.
\end{equation}
The four-dimensional Ricci scalar decomposes (up to total derivatives) as
\begin{equation}
R^{(4)} = R - \frac{2}{X}\,\nabla^{2}X + \frac{1}{2X^{2}}(\nabla X)^{2} + \frac{2}{X},
\end{equation}
where all derivatives refer to the two-dimensional metric \(g_{ab}\).
Integrating over the angular coordinates, $\int d\Omega_{2}=4\pi$, yields
\begin{equation}\label{eq:act2}
S_{2}=\frac{1}{4G_{4}}\int d^{2}x\,\sqrt{-g}\,
\Big[X R+U(X)(\nabla X)^{2}+2 V(X)\Big],
\end{equation}
with~\footnote{The factor of \(1/2\) in the definition of \(w'(X)\) is a convention; some SRG conventions instead absorb this factor into \(V\) or \(w\).}
\begin{equation}
U(X)=\frac{1}{2X},\qquad V(X)=1-\Lambda X.
\label{eq:UV_SRG_fixed}
\end{equation}

\subsection{SRG Casimir form and SdS solution}

It is useful to introduce the standard SRG integrating factor
\begin{equation}
Q'(X)=-U(X)=-\frac{1}{2X},
\end{equation}
so that $Q(X)=-\tfrac{1}{2}\ln X$ and $e^{Q(X)}=X^{-1/2}$.
We also define
\begin{equation}
\tilde V(X):=e^{Q(X)} V(X)
   =X^{-1/2}\!\left(1-\Lambda X\right).
\end{equation}
The functions $Q$, $\tilde V$, and $w$ define the SRG Casimir form of the
static solution.
Introducing
\begin{equation}\label{eq:wprime}
w'(X)=\frac{1}{2}\,\tilde V(X),\qquad
\xi(X)=e^{Q(X)}\big[w(X)-2G_4 M\big],
\end{equation}
the Schwarzschild--de~Sitter solution may be written in areal-radius coordinates as
\begin{equation}\label{eq:smetric}
ds^{2}=-\xi(r)\,dt^{2}+\xi(r)^{-1}dr^{2},\qquad X=r^{2}.
\end{equation}

Integration of \eqref{eq:wprime} gives
\begin{equation}
\begin{split}
w(X)
 &= \int^{X}\frac{1}{2}\tilde V(y)\,dy
   =\int \frac{1}{2}\Big(\tfrac{1}{r}-\Lambda r\Big)2r\,dr \\
 &= r-\tfrac{\Lambda}{3}r^{3}+{\rm const}.
\end{split}
\end{equation}

Choosing the constant to vanish and using $e^{Q}=1/r$, the metric function becomes
\begin{align}\label{eq:sdsmet}
\xi(r)
&=\frac{1}{r}\big(w-2G_{4}M\big)
 =1-\frac{2G_{4}M}{r}-\frac{\Lambda}{3}r^{2},
\end{align}
which reproduces the familiar Schwarzschild--de~Sitter blackening function.

\section{Semiclassical trace--anomaly backreaction}
\label{sec:framework}

We work in the large-$N$ semiclassical expansion of our spherically reduced gravity,
in which the backreaction of $N$ conformal matter fields is encoded by the
two-dimensional Polyakov anomaly action.
The full 2D effective action is
\begin{equation}
S_{\rm eff} = S_{2} + S_{\rm anom} \,,
\end{equation}
where $S_2$ is given by \eqref{eq:act2} with \eqref{eq:UV_SRG_fixed}, and the Polyakov term captures the trace anomaly of $N$ massless conformal fields,
\begin{equation}
S_{\rm anom}[g]
   = -\,\frac{N}{96\pi}\int d^2x\,\sqrt{-g}\,R\,\Box^{-1}R,
\end{equation}
whose variation yields the anomalous trace
$\langle T^\mu{}_\mu\rangle = (N/24\pi)\,R$.
For practical computations we employ the equivalent local (auxiliary-field)
form of $S_{\rm anom}$ given in Appendix~\ref{app:quadratures}.

\medskip
In conformal coordinates $(u,v)$ with $ds^2=-e^{2\rho(u,v)}du\,dv$,
the renormalized expectation values of the stress tensor are
\begin{align}\label{eq:stressall}
\langle T_{uu}\rangle &= -\frac{N}{12\pi}\big[(\partial_u\rho)^2-\partial_u^2\rho-t_u(u)\big], \nonumber \\
\langle T_{vv}\rangle &= -\frac{N}{12\pi}\big[(\partial_v\rho)^2-\partial_v^2\rho-t_v(v)\big], \nonumber \\
\langle T_{uv}\rangle &= -\frac{N}{12\pi}\,\partial_u\partial_v\rho,
\end{align}
where $t_u$ and $t_v$ encode the quantum state (e.g. Unruh–dS thermal values).
Regularity on the appropriate future Kruskal horizons fixes these constants:
for the Unruh--de~Sitter state, the future black-hole and cosmological horizons
are regular, producing a steady flux between them.

Note that in contrast to near-Nariai JT gravity analyses, where an explicit closed-form solution for the dilaton evaluated in the Unruh--de Sitter state is required in order to compute the generalized entropy exactly, our analysis does not rely on an explicit UdS--state dilaton profile. Instead, we work directly with the spherically reduced dilaton gravity model describing the full Schwarzschild–de Sitter static patch and determine the semiclassical geometry and stress tensor in the UdS state via the anomaly-induced effective action and Killing energy conservation. This is sufficient to fix the stationary Hawking flux, local thermodynamic observables, and entropy production without requiring an explicit analytic expression for the dilaton as a function of spacetime coordinates in the UdS state.

Combining the classical and anomaly actions yields a closed,
covariant system describing Hawking radiation, backreaction,
and mass evolution in the Schwarzschild--de~Sitter spacetime.
In what follows, we develop this framework to derive the conserved energy flux, the covariant mass–loss law, and the nonequilibrium thermodynamics of the system.

\subsection{Conserved flux and mass--loss law}
\label{sec:fluxlaw}

The steady Killing energy current admits a simple covariant formulation
that directly yields the semiclassical mass--loss law.
In the ingoing Eddington--Finkelstein (EF) gauge we have flux
\begin{equation}
\mathcal{J} := -\,I(X)\,X'(r)\,T^{r}{}_{v},
\qquad I(X)=e^{-Q(X)},\quad Q'(X)=-U(X),
\end{equation}
which is radially constant (and, to leading adiabatic order, slowly varying in~$v$).
For the Schwarzschild--de~Sitter reduction $X=r^{2}$ and $I(X)=\sqrt{X}=r$, giving
\begin{equation}
\mathcal J = -\,2r^{2}\,T^{r}{}_{v}.
\end{equation}
The conserved Killing energy current is $j^\mu=-T^\mu{}_\nu\chi^\nu$ with $\chi=\partial_t$.
In ingoing EF coordinates one has $\chi=\partial_v$, so the radial Killing flux is
$j^r=-T^r{}_v$, yielding
\begin{equation}
T^{r}{}_{v}
   =-\frac{\mathcal J}{I X'}
   =-\frac{\mathcal J}{2r^2}.
\end{equation}

\medskip
The EF Casimir (mass--function) balance law yields the covariant
mass--loss relation
\begin{equation}
\dot M = -\,\mathcal{J}.
\label{eq:massloss}
\end{equation}
Here $M$ denotes the mass parameter appearing in the Schwarzschild--de~Sitter
metric; equivalently, in the reduced two--dimensional description the Casimir is
$\mathcal{C}=2G_4 M$ (up to an additive constant).

The detailed radial first integrals underlying this relation appear in
Appendix~\ref{app:quadratures}.
Because the field equations do not admit a strictly static solution with
$\mathcal{J}\neq0$ (as proved in Appendix~\ref{app:quadratures}),
we work in the adiabatic steady--state regime, where $M$ evolves slowly and
the geometry remains quasi--static.

\begin{figure}[t]
\centering
\includegraphics[width=14cm]{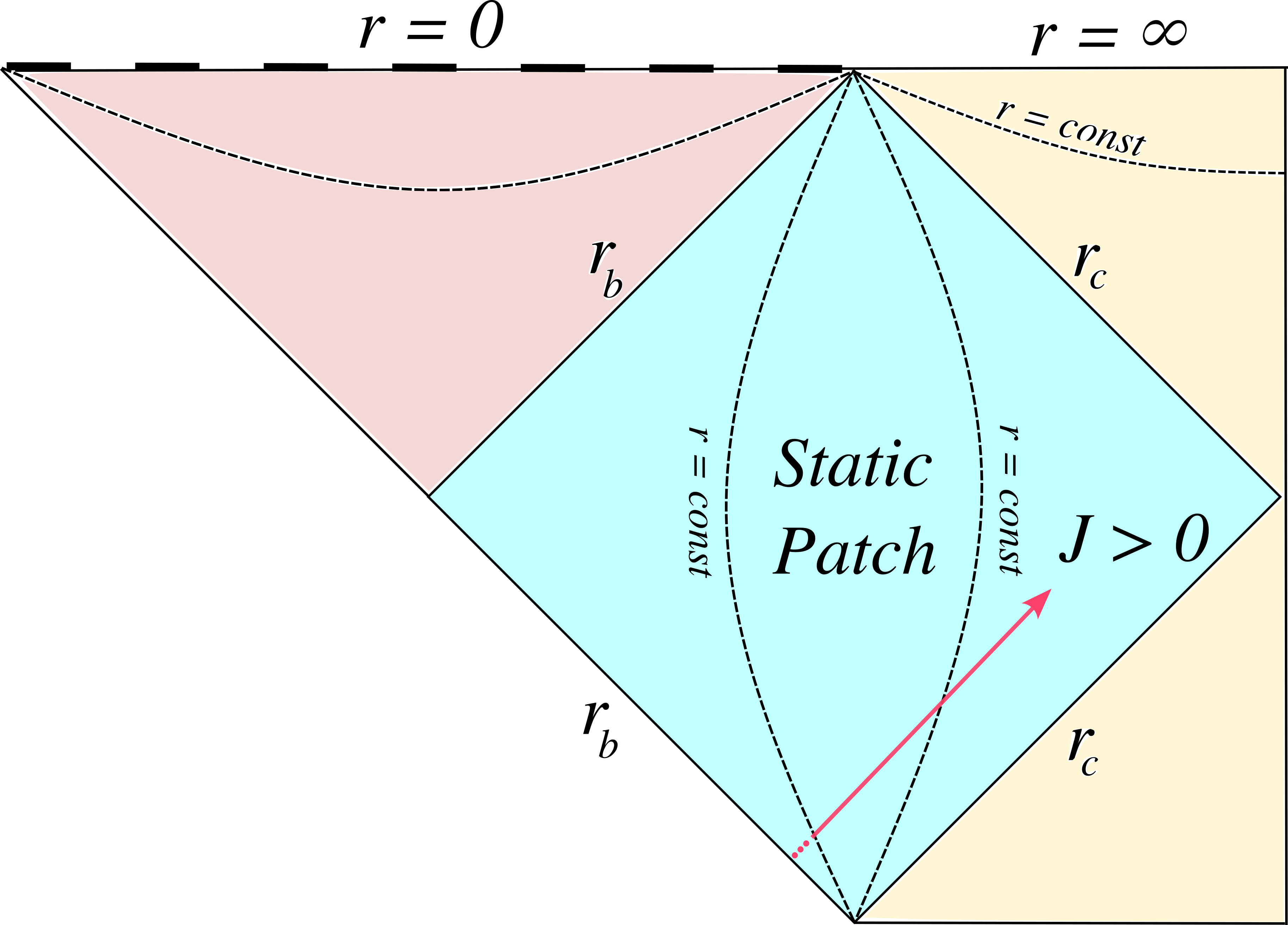}
\caption{\justifying
Portion of the Penrose diagram for Schwarzschild--de~Sitter
emphasizing the static patch (blue). The patch is bounded by
the black--hole ($r_b$) and cosmological ($r_c$) horizons. The black--hole singularity is at $r=0$ (horizontal dashed line). 
Dashed curves represent surfaces of constant $r$. 
In the Unruh--de~Sitter state, the anomaly--induced flux 
$\mathcal{J}>0$ propagates outward from the hotter black--hole horizon to the 
cooler cosmological horizon, establishing a steady nonequilibrium state.}
\label{fig:staticpatch}
\end{figure}

\medskip
For the Schwarzschild--de~Sitter metric
the surface gravities at the horizons $r_h\in\{r_b,r_c\}$ are
$\kappa_h=\tfrac12|\xi'(r_h)|$.  
The geometry contains a finite static patch bounded by the black--hole and cosmological horizons (see Fig.~\ref{fig:staticpatch}), representing the causal domain of a stationary observer.

\medskip
In the Unruh--de~Sitter state, the anomaly--induced stress tensor admits a steady flux
\begin{equation}
\mathcal{J}=\frac{N}{48\pi}\,\Delta(M),
\qquad 
\Delta(M)\equiv\kappa_b^2-\kappa_c^2,
\label{eq:Deltadef}
\end{equation}
directed outward for $\kappa_b>\kappa_c$.  
Identifying $M$ with the Casimir mass parameter controlling the SdS geometry,
the evolution follows

\begin{equation}\label{eq:fluxout}
\dot M=-\,\frac{N}{48\pi}\,\left(\kappa_b^2-\kappa_c^2 \right),
\end{equation}
so $\dot M<0$.  

Linearizing near a fixed point $M_\ast$ gives
\begin{equation}
\frac{d}{dt}\delta M
= -\frac{N}{48\pi}\,\Delta'(M_\ast)\,\delta M,
\label{eq:Linm}
\end{equation}
which is stable if $\Delta'(M_\ast)>0$ and unstable if $\Delta'(M_\ast)<0$.  
For neutral SdS, $\kappa_b=\kappa_c$ occurs only in the degenerate Nariai limit; thus no finite--temperature equilibrium exists within the static patch.

\medskip
We assume the evaporation is slow ($|\dot{M}|\ll \kappa_b M$), so that at each instant
the geometry remains approximately static (adiabatic approximation).
This justifies treating the Hawking flux as spatially and temporally constant during the
slow evaporation of $M$.\footnote{This approximation is well justified for large, slowly
radiating black holes.  As $M$ approaches the Planck scale or the minimal mass allowed by
the effective theory, the semiclassical and adiabatic descriptions necessarily break down
and quantum--gravity effects are expected to dominate the final stage of evaporation.}

\medskip
The locally measured manifestation of this flux, together with the Tolman blueshift
experienced by static observers, is analyzed in Sec.~\ref{sec:localFlux}.

\subsection{Local temperatures and Tolman redshift}
\label{sec:TolmanSetup}

In conformal null coordinates, $ds^2=-e^{2\rho}du\,dv$ (so that $e^{2\rho}=\xi$ in the static patch),
the Polyakov stress tensor involves two state functions $t_u(u)$ and $t_v(v)$ which determine
the flux components $\langle T_{uu}\rangle$ and $\langle T_{vv}\rangle$.
Regularity on a future horizon with surface gravity $\kappa_h$ fixes the corresponding function
to $t_{u/v}=\kappa_h^{2}/4$, i.e.\ a thermal population at temperature
\begin{equation}
T_h=\frac{\kappa_h}{2\pi}.
\label{eq:HawkingTemp}
\end{equation}
Different quantum states correspond to different choices of $(t_u,t_v)$:
\begin{itemize}
  \item \textbf{Boulware--dS:} $t_u=t_v=0$, static but singular on both horizons;
  \item \textbf{Hartle--Hawking--dS:} $t_u=t_v$ with $\kappa_b=\kappa_c$
        (global thermal equilibrium, realized only at the Nariai point);
  \item \textbf{Unruh--dS:} $t_u=\kappa_b^2/4$, $t_v=\kappa_c^2/4$,
        regular on the future horizons and describing steady outward flux.
\end{itemize}

Static observers at $r\!\in\!(r_b,r_c)$ measure locally redshifted temperatures according to the
Tolman--Ehrenfest relation $T_{\mathrm{loc}}\sqrt{-g_{tt}}=\text{const}$~\cite{Tolman:1930ona},
\begin{equation}
T_{h,\mathrm{loc}}(r)
   =\frac{T_h}{\sqrt{\xi(r)}}
   =\frac{\kappa_h}{2\pi\sqrt{\xi(r)}}.
\label{eq:TlocalConcept}
\end{equation}
The ratio $T_{b,\mathrm{loc}}/T_{c,\mathrm{loc}}=\kappa_b/\kappa_c$
is position-independent,
revealing the intrinsic nonequilibrium between the two horizons:
only in the degenerate Nariai limit do the local temperatures coincide everywhere.
This two-dimensional model exactly reproduces the four-dimensional
temperature hierarchy and equilibrium condition of
Schwarzschild--de~Sitter spacetime,
providing a faithful semiclassical reduction of the 4D thermodynamics.
The resulting steady state carries a constant energy flux between the horizons,
whose locally measured form is analyzed in Sec.~\ref{sec:localFlux}.

The quantities $T_h=\kappa_h/2\pi$ appearing in the First Law
\eqref{eq:firstlaw} are the standard \emph{Killing temperatures}, defined by
the surface gravities of the two horizons.  Local static observers, however,
measure Tolman-redshifted temperatures  
$T_{\rm loc}(r)=T_h/\sqrt{f(r)}$, which differ from $T_h$ except in special
equilibrium cases (e.g.\ the Nariai limit).  Because de~Sitter space lacks a
preferred asymptotic region, the distinction between Killing, Tolman, and
other proposed “global’’ temperature notions
(e.g.\ the one advocated in \cite{Volovik:2023stf}) is conceptually
important.

In the present work, this ambiguity does not affect the evaporation dynamics:
the mass-loss law is controlled entirely by the conserved Killing energy flux
$\mathcal{J}$, which is a radially invariant quantity and therefore
independent of which temperature convention one adopts.  The horizon Clausius
relations and the generalized second law follow directly from $\mathcal{J}$
and the geometric relation between $\kappa_b$ and $\kappa_c$, rather than from
any particular observer-adapted temperature definition.  For recent discussions of observer-dependent thermodynamics in multi-horizon
Reissner--Nordstr\"om--de~Sitter spacetimes, see~\cite{Aalsma:2025lcb}.

\section{Evaporation Dynamics and Mass Evolution}
\label{sec:evap}

From this point onward, unless explicitly restored, we work in Planck units
$G_4=\hbar=c=1$.

Using the flux \eq{eq:massloss}, we now analyze the time evolution of the system under the anomaly--induced backreaction.
The evolution of the mass parameter $M(t)$ follows from the interplay between the black-hole and cosmological surface gravities, which depend implicitly on $M$ and $\Lambda$ through the horizon equation $\xi(r_h;M,\Lambda)=0$.


The static-patch metric with \eq{eq:sdsmet},
admits two positive real roots $r_b<r_c$ when $0<9M^2\Lambda<1$, 
defining the black-hole and cosmological horizons, bounding the causal diamond of a static observer.  For fixed $\Lambda$, increasing $M$ enlarges $r_b$ and shrinks $r_c$, until they coincide at the Nariai limit $9M^2\Lambda=1$.


The surface gravities at the horizons are
\begin{equation}
\kappa_b=\frac{M}{r_b^2}-\frac{\Lambda}{3}r_b, 
\qquad
\kappa_c=\frac{\Lambda}{3}r_c-\frac{M}{r_c^2},
\label{eq:Surfgrav}
\end{equation}
with $\kappa_b,\kappa_c>0$ inside the static patch.  
Their difference determines the flux direction via $\Delta(M)$ (Eq.~\eqref{eq:Deltadef}).  

As the black hole evaporates ($M$ decreases), the black-hole horizon $r_b$ contracts while the cosmological horizon $r_c$ expands.
Consequently $\kappa_b$ rises rapidly, whereas $\kappa_c$ increases from its small Nariai value toward the asymptotic limit $\sqrt{\Lambda/3}$.
The growing temperature contrast $\Delta$ amplifies the Hawking flux, \eq{eq:Deltadef}.

\subsection{Flux, equilibrium and evolution timescales}

The qualitative behavior of the system is governed by sign of $\Delta(M)$:
\begin{align*}
\Delta>0 &:\quad \dot M<0 \;\;\Rightarrow\;\; \text{evaporation (energy flow $b\!\to\!c$)},\\
\Delta=0 &:\quad \dot M=0 \;\;\Rightarrow\;\; \text{thermal equilibrium},\\
\Delta<0 &:\quad \dot M>0 \;\;\Rightarrow\;\; \text{anti--evaporation}.
\end{align*}
For any nondegenerate interior equilibrium \(M_\ast\), as may occur in charged
or modified extensions, linearizing the flux law gives
\eqref{eq:Linm}
so the fixed point is locally attractive if \(\Delta'(M_\ast)>0\) and
repulsive if \(\Delta'(M_\ast)<0\). Neutral Schwarzschild--de~Sitter,
however, has no such interior equilibrium. Its only zero-flux point is the
degenerate Nariai boundary, where the two horizons coincide. Perturbations into
the physical static-patch domain have \(\Delta>0\), so the adiabatic flow
decreases \(M\) and moves the solution away from the Nariai boundary.

The adiabatic evaporation timescale follows from Eq.~\eqref{eq:massloss}:
\begin{equation}
\tau_{\rm evap}\sim
\frac{M}{|\dot M|}
\simeq\frac{48\pi M}{N|\Delta(M)|}.
\end{equation}
Evaporation is slow when the temperature contrast $|\Delta|^{1/2}$ is small and accelerates as $\kappa_b^2\!\gg\!\kappa_c^2$.  
For a nondegenerate interior equilibrium in a charged or modified extension,
the corresponding relaxation timescale is
\[
\tau_\ast \simeq \frac{48\pi}{N\,|\Delta'(M_\ast)|}.
\]
This estimate does not apply to the neutral Nariai boundary, which is a
degenerate endpoint of the static-patch domain rather than an ordinary
interior fixed point.

As $M\!\to\!0$, one finds $r_b\simeq 2M$, $\kappa_b\simeq(4M)^{-1}$, and $\kappa_c\simeq\sqrt{\Lambda/3}$, giving
\[
\Delta\simeq\frac{1}{16M^2}-\frac{\Lambda}{3},
\qquad
\dot M\simeq-\frac{N}{768\pi}\,M^{-2}.
\]
Hence $\tau_{\rm evap}\sim768\pi\,M^3/N$, reproducing the familiar $M^3$ scaling of four-dimensional Schwarzschild evaporation.


In principle, while three generic regimes exist:
\begin{enumerate}
\item \textbf{Monotonic evaporation:} $\Delta>0$ everywhere $\Rightarrow$ $M$ decreases to zero (neutral SdS).
\item \textbf{Stable equilibrium:} $\Delta(M)$ crosses zero with $\Delta'(M_\ast)>0$.
\item \textbf{Unstable equilibrium:} $\Delta(M)$ crosses zero with $\Delta'(M_\ast)<0$ ,
\end{enumerate}
only the first regime occurs in the neutral black hole case. Charged (Reissner--Nordstr\"om--de~Sitter) black holes or modified-gravity
extensions are known to admit interior solutions with
$\kappa_b=\kappa_c\neq0$ (``lukewarm'' configurations), and may therefore realize
the latter two regimes, but a detailed analysis of such cases lies beyond the
scope of this paper. 

We now analyze these equilibria in detail and identify the Nariai configuration as the unique zero-flux limit.

\section{Quasi-stationary geometry and localized Polyakov sector}
\label{sec:exactsteady}

In a strictly static metric ansatz,
the off--diagonal $(tr)$ equation enforces $T^{r}{}_{t}=0$,
so any exact static configuration must have vanishing flux ($J=0$).
The nonzero Unruh--de~Sitter flux $\mathcal{J}$ (the conserved
energy current in the quasi--stationary state) therefore arises consistently only
when time dependence is retained in Eddington--Finkelstein gauge,
or within a quasi--stationary (adiabatic) approximation in which
$M(v)$ evolves slowly compared to the local curvature scale.
Throughout this section we use the term ``steady state'' in this adiabatic sense.

The gravitational sector remains the original spherically reduced action
\eqref{eq:act2}. The functions $Q(X)$, $I(X)$, $\tilde V(X)$, and $w(X)$
are used as the standard SRG integrating-factor and Casimir machinery for
organizing the solution, rather than as a replacement of the SRG equations by
a pure $\tilde V$ model.
In this analysis, all leading quantum effects are incorporated through a
large-$N$, one-loop treatment of $N$ conformal fields.

\subsection{Localized Polyakov sector and semiclassical data}

The anomaly-induced Polyakov action can be written in local form by introducing an auxiliary field $\psi$:
\begin{equation}
S_{\rm P}=-\frac{N}{96\pi}\!\int d^2x\,\sqrt{-g}\,\left[(\nabla\psi)^2+2\psi\,R\right],
\end{equation}
whose equation of motion $\Box\psi=R$ reproduces the trace anomaly.
It is this localized representation of the Polyakov term that we use together
with the SRG action \eqref{eq:act2}; no separate pure-$\tilde V$ gravitational
action is assumed.

Near any simple horizon where $f(r)\simeq\kappa(r-r_h)$, the local
auxiliary-field solution has $f\psi'=-f'+C_\psi$.
The integration constant $C_\psi$, together with the state functions $t_\pm$,
is fixed by requiring the renormalized stress tensor to be regular in the
appropriate future Kruskal frames. Its explicit value is not needed for the
flux-balance argument.
In the strictly static ansatz this implies $T^{r}{}_{t}=0$, while in the adiabatic
Eddington--Finkelstein solution the physical flux is $\mathcal J$, related to the
static-frame component \eq{eq:Floc}.

\subsection{Solution structure and horizon regularity}

The quasistationary solution is characterized by the Casimir mass and by the
conserved SRG flux $\mathcal J$. Regularity of the renormalized stress tensor
on the appropriate future Kruskal horizons (black hole and cosmological)
imposes state-dependent boundary conditions consistent with the
Unruh--de~Sitter state. These conditions fix the state functions $t_\pm$,
equivalently the conserved EF flux $\mathcal J$, yielding an adiabatic
(quasi-stationary) solution at leading order in slow time.\footnote{In conformal gauge, the renormalized stress tensor depends on
two state functions $t_\pm$ that encode the choice of quantum state
(Boulware, Hartle--Hawking, or Unruh).  In the convention of Eq.~\eqref{eq:stressall},
the Unruh--de~Sitter choice $t_+=\kappa_b^2/4$ and $t_-=\kappa_c^2/4$
gives $\mathcal J=N(\kappa_b^2-\kappa_c^2)/(48\pi)$, so specifying the
flux $\mathcal J$ is equivalent to fixing $(t_+,t_-)$.}
The instantaneous geometry is parametrized by the Casimir (mass) and flux,
e.g.\ $(M,\Lambda,\mathcal J)$, and continuously deforms the classical SdS
spacetime while maintaining a radially conserved energy flux across the static
patch. The static no-flux constraint and the mass-loss balance are summarized
in Appendices~\ref{app:quadratures} and~\ref{app:EF}, respectively, and the
finiteness of the backreacted curvature invariants is discussed in
Appendix~\ref{app:curvature}.

This adiabatic setup provides a self-consistent,
analytic basis for generalized-entropy and extremal-surface calculations: given
$\langle T_{\mu\nu}\rangle$ and the Polyakov field, we compute
$S_{\rm gen}=S_{\rm Wald}(X)+S_{\rm out}(\mathcal R\cup I)$ in the Unruh--de~Sitter state,
locate QES via $\partial_\pm S_{\rm gen}=0$, and estimate Page-like behavior in a
two-horizon setting at leading semiclassical order.

This same anomaly--induced backreaction that governs the semiclassical dynamics of the SdS patch also plays a decisive role in other multi-horizon systems.
An analogous mechanism destabilizes the inner (Cauchy) horizons of charged and rotating black holes, where the trace anomaly is the minimal quantum enforcer of strong cosmic censorship by converting inner horizons into null curvature singularities~\cite{Birrell:1978th,Poisson:1990eh,Ori:1991zz,Brady:1995ni,Hollands:2019whz,Hollands:2020qpe,Cardoso:2018nvb,Dafermos:2003wr,Easson:2025uca}.

\section{Equilibrium Configurations and Stability}
\label{sec:equiconf}

Evaporation halts when the net Hawking flux between the black-hole and cosmological horizons vanishes. 
As derived in Eq.~\eqref{eq:Deltadef}, this occurs when the two surface gravities coincide,
\begin{equation}
\kappa_b(M_\ast) = \kappa_c(M_\ast).
\label{eq:equal_kappa_condition}
\end{equation}
Such equal-temperature configurations define equilibrium points $M_\ast$ for fixed $\Lambda$. 
In charged (Reissner--Nordstr\"om--de~Sitter) spacetimes, these correspond to the well-known \emph{lukewarm} black holes with $\kappa_b=\kappa_c\neq0$.
For the uncharged Schwarzschild--de~Sitter case considered here, however, this equality is achieved only at the degenerate \emph{Nariai} limit ($9M^2\Lambda=1$), where the two horizons merge ($r_b=r_c$) and both surface gravities vanish. 
Hence, the Nariai geometry represents the unique zero-flux equilibrium configuration of the neutral SdS family.

To discover the sign of the flux away from equilibrium, we derive an explicit analytic expression for $\Delta=\kappa_b^2-\kappa_c^2$. 
Eliminating $M$ from the two horizon equations,
\[
M=\tfrac{r_b}{2}\!\left(1-\tfrac{\Lambda}{3}r_b^2\right)
=\tfrac{r_c}{2}\!\left(1-\tfrac{\Lambda}{3}r_c^2\right),
\]
gives the fundamental horizon relation
\begin{equation}\label{eq:posdel}
1=\frac{\Lambda}{3}\, \left(r_b^2+r_b r_c+r_c^2 \right).
\end{equation}
Using this identity, the surface gravities satisfy
\begin{equation}
\kappa_b + \kappa_c
= \frac{r_c - r_b}{2}\!\left(\Lambda + \frac{1}{r_b r_c}\right) > 0,
\end{equation}
and, independently,
\begin{equation}
\kappa_b - \kappa_c
= \frac{r_b + r_c}{2}\!\left(\frac{1}{r_b r_c} - \Lambda\right) > 0,
\end{equation}
so that $\kappa_b>\kappa_c$ for all $0<9M^2\Lambda<1$.

\medskip

Hence,
\[
\Delta(M)=(\kappa_b-\kappa_c)(\kappa_b+\kappa_c)>0,
\]
and (as one may have anticipated) the semiclassical flux 
\[
\mathcal{J}=\tfrac{N}{48\pi}\,\Delta(M)
\]
is positive throughout the static patch, implying strictly monotonic mass loss ($\dot M<0$) for all nonextremal configurations.
Thus no finite-temperature interior equilibrium exists in neutral SdS: the only
zero-flux configuration is the degenerate Nariai boundary, which we discuss next.

\subsection{The Nariai limit and zero--flux equilibrium}
\label{sec:Nariai}

We have shown the SdS family admits a single equilibrium configuration where the black hole and cosmological horizons coincide.  This extremal geometry, known as the \emph{Nariai spacetime} \cite{Nariai:1999iok, Bousso:1997wi, Lemos:2024sjs}, arises at the maximal mass
\begin{equation}
M_{\rm Nariai} \;=\; \frac{1}{3\sqrt{\Lambda}},
\qquad
r_b = r_c = \frac{1}{\sqrt{\Lambda}},
\end{equation}
for which the surface gravities vanish $\kappa_b = \kappa_c = 0$.
In this limit the SdS coordinates become degenerate, but after a suitable rescaling the metric factorizes as
\begin{equation}
ds^2 = \frac{1}{\Lambda}
\!\left[-\,\sin^2\!\chi\, d\tau^2 + d\chi^2\right]
+ \frac{1}{\Lambda}\, d\Omega_2^2,
\end{equation}
describing the direct product $\mathrm{dS}_2\times S^2$; the two-dimensional
de~Sitter factor has scalar curvature \(R^{(2)}=2\Lambda\), while the full
four-dimensional product has \(R^{(4)}=4\Lambda\).
The Nariai geometry therefore represents a zero-temperature equilibrium
configuration of the SdS family in the original static normalization:
$\kappa_b=\kappa_c=0$ and the Killing flux vanishes. After the near-horizon
Nariai scaling, the limiting $\mathrm{dS}_2\times S^2$ geometry has its own
finite temperature with respect to the rescaled Nariai time.
Perturbations away from this limit split the horizons and produce nearby SdS solutions with $\kappa_b > \kappa_c$, initiating nonequilibrium energy flow from the black hole to the cosmological horizon.

In our dilaton gravity description, the Nariai state is realized when the effective potential satisfies
\begin{equation}
\tilde V(X_N)=0, \qquad X_N=\frac{1}{\Lambda},
\end{equation}
so that $w'(X_N)=0$ and the Killing function $\xi(X)$ develops a double root.
The anomaly--induced flux, \eq{eq:Deltadef}, then vanishes $\mathcal{J} =0$,
confirming that the Nariai limit marks the sole equilibrium point in the $(M,\Lambda)$ phase diagram.

For the observed cosmological constant, $\Lambda \simeq 10^{-52}\,\mathrm{m^{-2}}$, the Nariai horizon radius is $r_{\rm Nariai}=1/\sqrt{\Lambda}\!\sim\!10^{26}\,\mathrm{m}$, and the corresponding mass is $M_{\rm Nariai} \simeq c^2/(3G_4\sqrt{\Lambda})\!\approx\!2.3\times10^{22}\,M_\odot$. This mass is cosmologically large, comparable to the total mass contained within the observable universe, and represents the unstable upper mass bound for a black hole in the Schwarzschild--de~Sitter family. All known black holes (from stellar to supermassive, $\sim10$--$10^{10}\,M_\odot$) lie deep in the $M\ll M_{\rm Nariai}$ regime. As the semiclassical flux law (Eq.~\ref{eq:fluxout}) is shown in Appendix~\ref{app:EF} to remain evaporative ($\dot M<0$) for this entire physical range, the Nariai configuration is never approached dynamically. Consequently, the only physically attainable endpoint of SdS black-hole evaporation in our de~Sitter background is the empty de~Sitter state.

We now map these regimes across $(M,\Lambda)$ parameter space and identify the Nariai line $9M^2\Lambda=1$ as the boundary separating the evaporation domain from the degenerate equilibrium configuration.

\subsection{Phase diagram}
\label{sec:PhaseDiagram}

The equilibrium set $\Delta(M,\Lambda)=0$ reduces, for neutral SdS, to the Nariai curve $9M^2\Lambda=1$, which forms the boundary of the physical static-patch domain. Within the domain $0<9M^2\Lambda<1$ one has $\Delta>0$, so the flux direction does not change sign.

\begin{figure}[t]
\centering
\includegraphics[width=14cm]{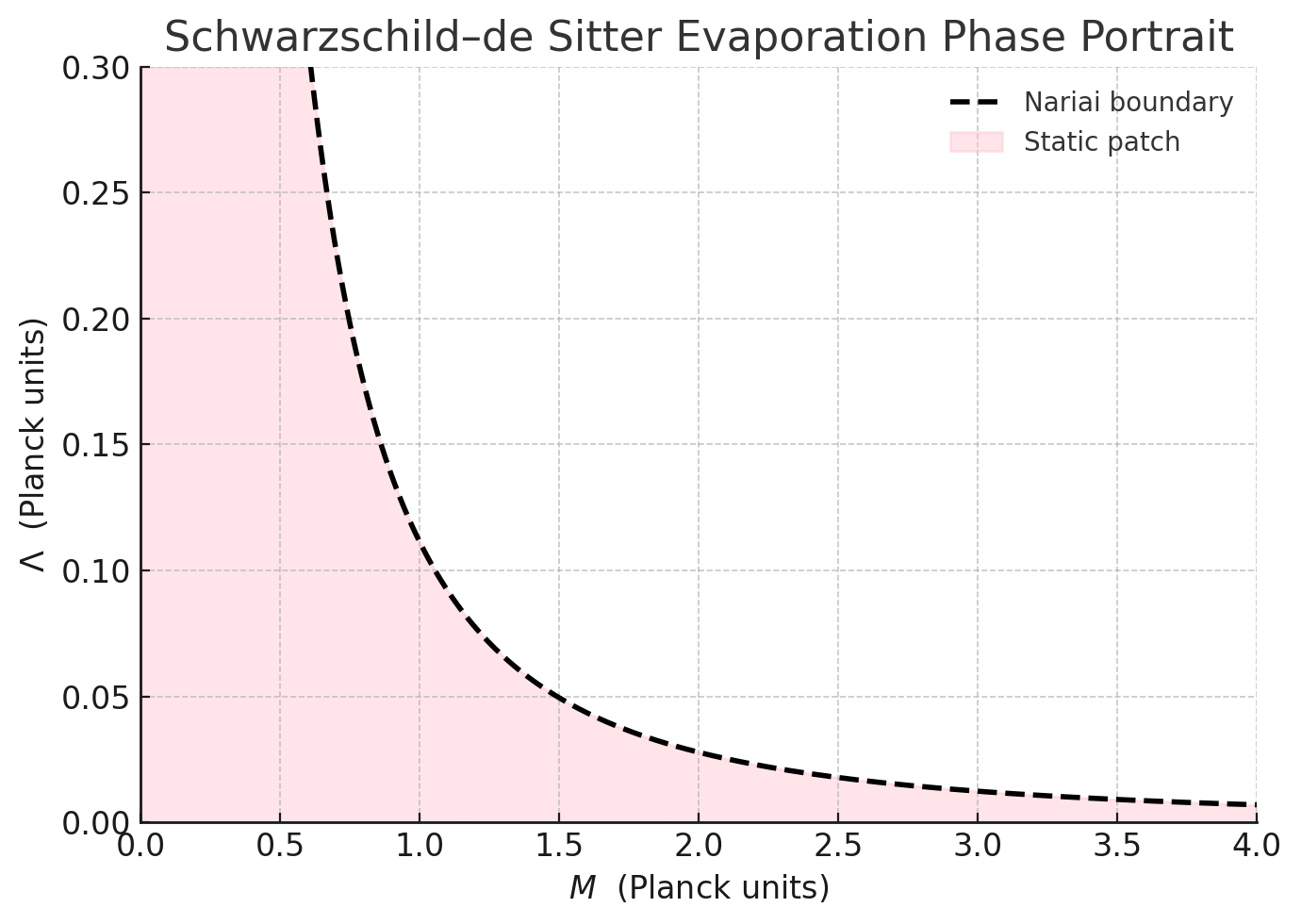}
\caption{\justifying
Evaporation phase diagram for Schwarzschild--de~Sitter black holes.
Each point $(M,\Lambda)$ with $0<9M^2\Lambda<1$ represents a physical static patch.
Throughout this domain $\Delta>0$, corresponding to a
steady energy flux from the black-hole to the cosmological horizon
($\dot M=-\mathcal J<0$).  The dashed curve marks the Nariai boundary
$9M^2\Lambda=1$, where the two horizons coincide and attain equal surface
gravities (see Sec.~VI~A).  All quantities are in Planck units
($G_4=\hbar=c=1$).
}

\label{phase}
\end{figure}

The Nariai curve is a degenerate boundary of the physical static-patch domain,
rather than an ordinary interior equilibrium line. Perturbations into the
physical domain \(0<9M^2\Lambda<1\) have \(\Delta>0\), so the anomaly-induced
flow decreases \(M\) and carries the solution away from the Nariai boundary.
This defines the semiclassical phase diagram of SdS evaporation shown in
Fig.~\ref{phase}.

The values of \(\Lambda\) shown are far larger than the observed cosmological
constant, but the analysis cleanly isolates the qualitative structure of the
evaporation dynamics. No nondegenerate equilibrium configuration exists within
the static patch: the anomaly-induced flux drives the black-hole mass
monotonically toward \(M\to0\), corresponding to empty de~Sitter, while the
Nariai boundary marks the unique zero-flux limit separating the evaporating
domain from the unphysical region of parameter space.

Having established the global structure of evaporation and the location of equilibrium boundaries in $(M,\Lambda)$ space, we now turn to the thermodynamics experienced by local static observers within the de~Sitter static patch.

\section{Observer-dependent thermodynamics}\label{sec:odt}

In the static patch of SdS, stationary observers experience distinct local thermodynamic conditions depending on their position between the black hole and cosmological horizons.
The region $r_b<r<r_c$ admits the timelike Killing field $\chi^\mu=\partial_t$ with norm $\chi^2=g_{\mu\nu}\chi^\mu\chi^\nu=-\xi(r)$. 
Static observers follow orbits of $\chi^\mu$ with four-velocity
\begin{equation}
u^\mu=\frac{\chi^\mu}{\sqrt{-\chi^2}}=\frac{1}{\sqrt{\xi(r)}}(1,0)\,,
\qquad
e^\mu_{(\hat r)}=\sqrt{\xi(r)}\,(0,1)\,,
\end{equation}
defining the orthonormal frame $\{u^\mu,e^\mu_{(\hat r)}\}$.

\subsection{Local temperatures and steady flux}
\label{sec:localFlux}

Building on the Tolman relation introduced in Sec.~\ref{sec:TolmanSetup},
we now evaluate the locally measured energy flux and density experienced
by static observers in the Schwarzschild--de~Sitter static patch.

Each Killing horizon $r_h\in\{r_b,r_c\}$ possesses surface gravity
$\kappa_h=\tfrac12|\xi'(r_h)|$ and Hawking temperature $T_h=\kappa_h/(2\pi)$.
A static observer at radius~$r$ measures the redshifted temperature
\begin{equation}
T_{h,\mathrm{loc}}(r)
   = \frac{T_h}{\sqrt{\xi(r)}}
   = \frac{\kappa_h}{2\pi\sqrt{\xi(r)}}\,,
\label{eq:TlocalFinal}
\end{equation}
so that $T_{b,\mathrm{loc}}/T_{c,\mathrm{loc}}=\kappa_b/\kappa_c$.
The patch is therefore intrinsically out of equilibrium whenever $\kappa_b\neq\kappa_c$;
only in the Nariai (lukewarm) limit do the local temperatures coincide.

To quantify this nonequilibrium steady state, let
$\langle T_{\mu\nu}\rangle$ denote the renormalized stress tensor
in the chosen quantum state.
For static observers with four-velocity $u^{\mu}=(1/\!\sqrt{\xi},0)$
and orthonormal radial vector $e^{\mu}_{(\hat r)}=(0,\sqrt{\xi})$,
the locally measured energy density and radial flux are
\begin{align}
\rho(r) &= \langle T_{\mu\nu}\rangle u^{\mu}u^{\nu},
\nonumber\\[2pt]
\mathcal F_{\mathrm{loc}}(r)
   &= -\,\langle T_{\mu\nu}\rangle u^{\mu} e^{\nu}_{(\hat r)}
    = -\,\frac{T^{r}{}_{t}}{\xi(r)}.
\label{eq:rhoFluxLocal}
\end{align}
Stationarity implies conservation of the outward SRG Killing flux
$\mathcal J=-2r^{2}T^{r}{}_{t}$ throughout the patch. Thus
\begin{equation}
T^{r}{}_{t}=-\frac{\mathcal J}{2r^{2}}.
\label{eq:Jconst}
\end{equation}
Projecting onto the local frame gives
\begin{equation}
\mathcal F_{\mathrm{loc}}(r)
   = \frac{\mathcal J}{2r^{2}\xi(r)}\,,
\label{eq:Floc}
\end{equation}
which blueshifts as $1/\xi(r)$ toward either horizon.
Hence the SdS static region acts as a finite thermal cavity
bounded by two horizons at unequal Tolman temperatures,
linked by a conserved heat current $\mathcal J$
that vanishes only in the Nariai equilibrium configuration.

Physically, $\mathcal J>0$ corresponds to an outward flux of Killing energy
from the hotter black-hole horizon to the cooler cosmological horizon.
The same flux governs the global mass-loss law
$\dot M=-\mathcal J$ derived in Sec.~\ref{sec:fluxlaw},
ensuring that the locally observed nonequilibrium
and the global evaporation dynamics are two aspects
of the same conserved-energy flow.

\subsection{Near-horizon behavior and regularity}

Near each Killing horizon, it is convenient to express the metric in conformal gauge,
\[
ds^2=-e^{2\rho}du\,dv, \qquad e^{2\rho}=\xi(r).
\]
The anomaly-induced stress tensor then takes the form~\eqref{eq:stressall},
where $t_u$ and $t_v$ are the state-dependent constants specifying the ingoing and outgoing
flux components in a stationary patch.

Regularity on a given future horizon fixes the corresponding constant to its thermal value
at $T_h=\kappa_h/(2\pi)$:
$t_u$ at the future black-hole horizon and $t_v$ at the future cosmological horizon.
The \emph{Unruh--de~Sitter} state enforces regularity on both future horizons while remaining
singular on the past ones, yielding the steady flux~\eqref{eq:Deltadef}.
In contrast, the \emph{Boulware--de~Sitter} state has $t_u=t_v=0$ and is singular on all horizons,
while the \emph{Hartle--Hawking--de~Sitter} state occurs only at the Nariai limit~\cite{Numajiri:2024qgh}.

\subsection{Detector response}

A standard static Unruh--DeWitt detector following the worldline
\(x^\mu(\tau)\), with four-velocity \(u^\mu\), interacts locally with the
quantum field and can become excited by absorbing a quantum of energy
\(\Omega>0\).  The quantity \(\mathcal{P}(\Omega;r)\) denotes the corresponding
transition probability, and \(\dot{\mathcal{P}}(\Omega;r)\) its transition rate
per unit proper time.  At leading order in perturbation theory this rate is
given by
\begin{equation}
\dot{\mathcal{P}}(\Omega;r)
   \;=\;
   \int_{-\infty}^{+\infty}\!d\tau\;
   e^{-i\Omega\tau}\,
   G^{+}\!\big(x(\tau),x(0)\big),
\end{equation}
where \(G^{+}\) is the Wightman function in the chosen quantum state.

In the stationary patch, the detector response is governed by the same
state data \(t_\pm\) that determine the Polyakov stress tensor and the
conserved flux \(\mathcal J\).  In the Unruh--de~Sitter state, the response
receives contributions associated with the black-hole and cosmological
horizon temperatures, each redshifted by the Tolman factor
\(1/\sqrt{\xi(r)}\), as in Eq.~\eqref{eq:TlocalConcept}.  Away from equilibrium
there is no single KMS temperature characterizing the detector response; the
nonequilibrium bias is controlled by the same temperature contrast that fixes
the outward Killing-energy flux \(\mathcal J\).

In the equal-temperature limit, by contrast, the response satisfies the usual
KMS detailed-balance relation at the common Tolman temperature.  For a detector
gap \(\Omega>0\), this relation may be written schematically as
\begin{equation}
\frac{\dot{\mathcal P}(+\Omega;r)}
     {\dot{\mathcal P}(-\Omega;r)}
 =
 e^{-\Omega/T_{\rm loc}(r)} .
\end{equation}
Thus the excitation rate itself need not vanish in equilibrium.  What vanishes
there is the net nonequilibrium heat current, equivalently the conserved flux
\(\mathcal J\), which is zero at the Nariai boundary; see
Sec.~\ref{sec:Nariai}.\footnote{Static detectors have infinite proper
acceleration near the horizons.  The Tolman temperature diverges as
\(1/\sqrt{\xi}\), while the locally measured energy flux scales as \(1/\xi\).
This divergence is associated with the static observer frame; the regularity
condition for the quantum state is imposed in freely falling/Kruskal frames.}

\section{Entropy flow and the generalized second law}
\label{sec:GSL}

The coexistence of two horizons in SdS
raises subtle questions about how gravitational entropy evolves during
evaporation. Here we show that the anomaly–induced flux $\mathcal{J}$
automatically satisfies the generalized second law (GSL) throughout the
evaporation process, as expected. In a quasistationary process, the Hawking radiation
within the static patch forms a steady flow rather than an accumulating bath;
its entropy therefore remains constant in time, $\dot S_{\rm out}=0$. The only
entropy changes arise from the horizons themselves, so that
the generalized entropy evolves as
\begin{equation}
\dot S_{\rm gen}=\dot S_b+\dot S_c.
\end{equation}

\subsection{Entropy balance and first-law structure}

Let $S_b=\pi X_b/G_4$ and $S_c=\pi X_c/G_4$ denote the Wald entropies of the
black-hole and cosmological horizons ($X=r^2$).  
At fixed $\Lambda$, the static SdS family obeys the parametric identity
\begin{equation}
2\delta M = T_b\,\delta S_b - T_c\,\delta S_c,
\label{eq:firstlaw}
\end{equation}
where the minus sign reflects the opposite Killing orientation of the
cosmological horizon.\footnote{ 
Some authors define the Wald entropy as $S=2\pi X$, which leads to a 
first law of the form $\delta M = 2 T\,\delta S$, whereas in the 
spherical-reduction normalization used here 
$S = X/(4G_2)$ (equivalently $S=\pi r^2/G_4$) reproduces the standard 
4D area law and yields the normalization used in Eq.~\eqref{eq:firstlaw}.  
Similarly, the temperatures $T_h=\kappa_h/2\pi$ are the Killing 
temperatures that enter the geometric first law, while local static 
observers measure Tolman-redshifted values 
$T_{\rm loc}=T_h/\sqrt{f(r)}$.  
This distinction does not affect the mass-loss law, which depends only 
on the conserved Killing flux $\mathcal J$.}
  This ``first law'' describes variations among nearby
static configurations and should not be interpreted as a dynamical evolution
equation.  The time-dependent semiclassical evolution is governed instead by
the conserved energy flux.

\subsubsection{Flux balance and horizon Clausius relations}
Energy conservation implies a single
radially conserved flux of Killing energy across the patch.  Defining
$\mathcal{J}>0$ as the outward (increasing-$r$) flux measured by static
observers, each horizon obeys its own local Clausius relation,
\begin{equation}
T_b\,\dot S_b = -\,\mathcal{J},
\qquad
T_c\,\dot S_c = +\,\mathcal{J},
\label{eq:Clausius_local}
\end{equation}
expressing that the black hole loses heat while the cosmological horizon gains
the same amount.  
The net Killing-energy balance of the entire patch is
\begin{equation}
\dot M = -\,\mathcal{J},
\label{eq:Massloss_J}
\end{equation}
so that each relation in \eqref{eq:Clausius_local} may equivalently be written as
\[
\dot M = T_b\dot S_b = -\,T_c\dot S_c.
\]
Equation \eqref{eq:firstlaw} therefore serves as a geometric identity among neighboring static configurations, whereas \eqref{eq:Clausius_local}–\eqref{eq:Massloss_J} govern the true time-dependent evolution driven by the anomaly-induced flux.

\subsubsection{Generalized second law}
Combining \eqref{eq:Clausius_local} gives the total gravitational entropy
production rate,
\begin{equation}
\dot S_{\rm gen}
   = \dot S_b + \dot S_c
   = \mathcal{J}\!\left(\frac{1}{T_c}-\frac{1}{T_b}\right) > 0
   \quad (\kappa_b>\kappa_c),
\label{eq:GSLfixed}
\end{equation}
showing that after the decrease in black-hole entropy, the cosmological
horizon's increase more than compensates, ensuring the monotonic growth of the
total generalized entropy.  When $T_b=T_c$ (the Nariai limit),
$\mathcal{J}=0$ and $\dot S_{\rm gen}=0$, indicating exact thermal equilibrium.

For conformal matter, the anomaly-induced flux
$\mathcal{J}=(\pi N/12)(T_b^2-T_c^2)$ reproduces the steady-state relation of
a $1{+}1$-dimensional CFT, and the entropy production rate becomes
\begin{equation}
\dot S_{\rm gen}
 = \frac{\pi N}{12}\,
   \frac{(T_b-T_c)^2\,(T_b+T_c)}{T_b T_c}
 \ge 0,
\end{equation}
with equality only in the Nariai configuration.  
Thus, throughout the physical static-patch domain $0<9M^2\Lambda<1$, the
Unruh–de~Sitter state satisfies $\mathcal{J}>0$ and $\dot S_{\rm gen}>0$,
while in the degenerate Nariai limit (where the horizons coincide and the
temperature contrast vanishes) the flux and entropy production both go to
zero. These relations hold on the two-horizon patch domain.
In the formal endpoint $M\!\to\!0$, the black-hole horizon disappears and
$T_b$ is no longer defined; the steady two-horizon flux formula is therefore
inapplicable, and the state smoothly reduces to the de Sitter vacuum with
no inter-horizon flux.
The 2D anomaly framework provides a fully analytic
and self-consistent realization of the generalized second law for spacetimes
with multiple horizons.

\section{Quantum Extremal Surfaces and Entanglement Islands}
\label{sec:islands}

The SdS solution constructed above naturally lends itself to the study of quantum
extremal surfaces (QES) and entanglement islands in a cosmological setting.
Recent developments in semiclassical gravity have shown that these quantum surfaces (stationary points of the generalized entropy functional) play a central role in
the unitary description of black hole evaporation
\cite{Almheiri:2019psf,Penington:2019npb,Almheiri:2020cfm}.
While most existing analyses are carried out in asymptotically flat or AdS spacetimes,
the Schwarzschild--de~Sitter geometry provides a conceptually distinct testing ground
with two horizons and a finite static patch.

\subsection{Generalized entropy and QES conditions}

In our 2D effective theory, the generalized entropy of a region
$\mathcal{R}$ with boundary $\partial\mathcal{R}$ is
\begin{equation}
S_{\rm gen}[\partial\mathcal{R}]
  = \frac{X(\partial\mathcal{R})}{4G_2}
    + S_{\rm out}[\mathcal{R}],
\label{eq:Sgen_island}
\end{equation}
where the first term represents the Bekenstein--Hawking (dilaton) entropy of the boundary surface,
and $S_{\rm out}$ is the von~Neumann entropy of quantum fields in $\mathcal{R}$.
The two-dimensional Newton constant $G_2$ is defined so that
$S_{\rm BH}=X/4G_2$ reproduces the usual four-dimensional area law;
for spherical reduction one has $G_2 = G_4/(4\pi)$,
ensuring \(S_{\rm BH} = \pi X / G_4 = X / (4G_2)\).~\footnote{In four dimensions the Bekenstein--Hawking entropy is
$S=A/(4G_4)$, with $A=4\pi r_B^2$.  Under spherical reduction the dilaton
is defined by $X=r^2$, so that $A=4\pi X$ and the reduced coupling
$G_2=G_4/(4\pi)$ yields
$S_{\rm BH}=\pi X/G_4=X/(4G_2)$.
Thus $X$ plays the role of the horizon area in the two-dimensional theory.}

A quantum extremal surface (QES) is a stationary point of the generalized entropy,
\begin{equation}
\partial_{\pm} S_{\rm gen}
   = \frac{1}{4G_2}\,\partial_{\pm} X
   + \partial_{\pm} S_{\rm out} = 0,
\label{eq:QES_cond}
\end{equation}
where $\partial_{\pm}$ denote derivatives along the outgoing and ingoing null directions
$k_\pm^\mu$ of the metric $ds^2=-e^{2\rho(u,v)}du\,dv$
(with $\partial_\pm \equiv k_\pm^\mu \nabla_\mu$).
These equations express the vanishing of the ``quantum expansions'' of $S_{\rm gen}$
and reduce to the usual extremality of the area term when quantum effects are neglected.

For conformal matter, $S_{\rm out}$ can be computed directly from the conformal factor
and the state-dependent functions that determine $\langle T_{\mu\nu}\rangle$.
In a smooth metric $ds^2=-e^{2\rho}du\,dv$, the entropy of an interval with endpoints
$p_i=(u_i,v_i)$ is
\begin{equation}
S_{\rm out}(p_1,p_2)
  = \frac{c}{6}\!\left[
      \rho(p_1)+\rho(p_2)
      + \ln\!\frac{|u_1-u_2|\,|v_1-v_2|}{\epsilon_1\epsilon_2}
    \right]
    + \delta_{\rm state},
\label{eq:Sinterval}
\end{equation}
where $c=N$ is the central charge, $\epsilon_i$ are short-distance cutoffs, and $\delta_{\rm state}$ encodes
the conformal transformation to coordinates regular on the relevant future horizons.
In our setup, the state is characterized by the constants $(t_u,t_v)$ that fix the
Unruh--de~Sitter flux and determine both $\langle T_{\mu\nu}\rangle$ and the additive
terms in $S_{\rm out}$.\footnote{In the stationary Unruh--de~Sitter state, the local
energy and entropy densities are time-independent, but the coarse-grained entropy carried
by the steady radiation flow increases linearly with time at the rate
$\dot S_{\rm prod}=\mathcal{J}\!\left(\tfrac{1}{T_c}-\tfrac{1}{T_b}\right)$.
Below we use this quantity only as a thermodynamic entropy-production proxy for
the Page-time estimate.}

\subsection{Islands in the static patch}

In the SdS static patch, a natural entanglement region $\mathcal{R}$
is the domain accessible to a static observer at fixed $r<r_c$,
bounded by the cosmological horizon.
Radiation emitted from the black-hole horizon propagates outward and becomes causally
inaccessible to the static observer beyond $r_c$.  Our explicit anomaly-induced backreaction and steady flux framework provides a background to analyze the information flow.

As the coarse-grained entropy of the outgoing radiation increases,
a stationary point of the generalized entropy $S_{\rm gen}$ can develop just inside the black-hole horizon $r_b$, marking the formation of an island whose interior lies beyond the horizon but whose entanglement wedge is contained within the static patch accessible to the observer.\footnote{In the two-dimensional reduction, the quantum extremal surface is a single
spatial point $r_{\rm I}$ (or a pair of points on opposite sides of the Penrose
diagram), so the ``island'' corresponds to the interval between this point and
the observer’s region~$\mathcal R$.}

At this point the dominant saddle in the path integral is expected to shift from the no-
island configuration to one that includes the island. This transition represents the de Sitter
analogue of a Page transition. Because both the anomaly-induced stress tensor and the
Polyakov field are known analytically in our model, the framework is well suited to a more
complete QES treatment of Page-like behavior in multi-horizon spacetimes.

\subsubsection{Entanglement structure}
In the de~Sitter static patch, the radiation field is not entangled solely with the
black-hole interior (as in asymptotically flat evaporation) but also with degrees
of freedom associated with the cosmological horizon.  The global state of the full
SdS spacetime is pure, yet a static observer has access only to the inter-horizon
region.  The relevant entanglement structure is therefore tripartite,
$(\mathcal{H}_{\rm BH}\!\otimes\!\mathcal{H}_{\rm rad}\!\otimes\!\mathcal{H}_{\rm dS})$, where each $\mathcal H$ denotes the Hilbert space of the corresponding subsystem
and the entropy accessible within the static patch is expected to be bounded by
the black-hole entropy, up to correlations with the cosmological horizon. The
Page transition corresponds to the point at which an island forms near the
black-hole horizon, incorporating its interior into the entanglement wedge of
the radiation in the full fine-grained description.

Our Page-curve analysis employs a two-dimensional dilaton reduction in which gravity has no local propagating degrees of freedom; the only fields with long-range tails are the conformal matter sectors entering the Polyakov action. In this setting, the Gauss-law constraints and gravitational dressings that obstruct Hilbert-space factorization in theories with dynamical (massless) gravitons are absent. Although diffeomorphism constraints remain, there is no soft-graviton sector, so the standard island prescription can be applied without additional edge-mode bookkeeping. Consequently, our results should be read as evidence \emph{within this controlled model} that, when long-range dressings are under control, the island mechanism operates consistently in the de Sitter static patch. 

Recent work has emphasized conceptual issues for massless gravity and gauge theories,
where factorization fails due to soft modes and global constraints
(see, e.g., \cite{Geng:2023iqd, Harlow:2018tng, Donnelly:2016auv, Donnelly:2014fua, Chandrasekaran:2022cip}). Our 2D dilaton model sidesteps these issues by construction, providing a semiclassical test-bed in which the island mechanism is unambiguous. Extending the analysis to retain four-dimensional soft-graviton sectors (including edge modes at the cosmological horizon and their charges) is an interesting open problem that would more directly address concerns specific to massless gravity.\footnote{Some braneworld constructions realize islands alongside an effectively massive graviton on the brane, but islands do not \emph{require} massive gravity. In the present 2D dilaton model, there is no propagating graviton at all, and the island construction remains consistent.}

\subsubsection{Cosmological counterpart}
By symmetry, a complementary quantum extremal surface (QES) can lie just inside
the cosmological horizon $r_c$, corresponding to an island for observers whose
algebra is anchored outside the static patch, in the neighboring de~Sitter region.
In the \emph{Unruh--de~Sitter} state considered here, the steady flux is directed
outward from the black--hole horizon toward the cosmological horizon.
For observers within the static patch, this makes the inner island near~$r_b$
the dynamically relevant saddle, while a cosmological--side island would instead
be relevant to observers anchored beyond~$r_c$.
In time--reversed Unruh states the roles of the horizons interchange.

While cosmological- and black-hole–side quantum extremal surfaces have been identified in equilibrium and near-horizon models, the role of a conserved, anomaly-induced nonequilibrium flux in selecting the physically relevant island saddle for a given observer in the full SdS static patch has not been explicitly analyzed previously.

In true equal--temperature (\emph{lukewarm}) equilibrium there is no net flux:
symmetric island saddles may then exist, for instance in charged
Reissner--Nordstr\"om--de~Sitter families.
For the neutral Schwarzschild--de~Sitter geometry, however, equality of the
horizon temperatures occurs only in the degenerate Nariai limit.
In all cases, candidate island endpoints satisfy the extremality conditions
$\partial_\pm S_{\rm gen}=0$; which saddle dominates and the onset of the
island phase (Page time) depend on the choice of quantum state and the flux
orientation, and will be quantified below.

\subsection{A thermo-controlled Page-curve proxy}
\label{sec:page-curve}

We refer to this construction as \emph{thermo-controlled} because the
growth scale and estimated turnover are determined entirely by macroscopic
thermodynamic quantities: the horizon temperatures and steady flux. The
quantity below is therefore a coarse-grained thermodynamic proxy rather than a
microscopic fine-grained entropy derived from an explicit QES calculation.

As we have determined, in the Unruh--de~Sitter state with $\kappa_b>\kappa_c$, the static patch supports a steady
outward Killing flux
\begin{equation}
\mathcal J
 = \frac{N}{48\pi}\big(\kappa_b^2-\kappa_c^2\big)
 = \frac{N\pi}{12}\big(T_b^2-T_c^2\big),
\qquad
T_h=\frac{\kappa_h}{2\pi}.
\label{eq:J-Ts}
\end{equation}
The corresponding irreversible (local) entropy production rate
(Sec.~\ref{sec:GSL}) is
\begin{equation}
\dot S_{\rm prod}
 = \mathcal J\!\left(\frac{1}{T_c}-\frac{1}{T_b}\right)
 \;>\; 0
 \qquad (\kappa_b>\kappa_c\ \text{or}\ T_b>T_c),
\label{eq:Sprod}
\end{equation}
representing a steady flow of heat and entropy from the black--hole to the cosmological horizon.

To leading adiabatic order, this rate defines the coarse-grained
thermodynamic entropy-production proxy
\begin{equation}
\dot S_{\rm proxy}(t)
  \simeq
  \mathcal J\!\left(\frac{1}{T_c}-\frac{1}{T_b}\right),
\qquad
S_{\rm proxy}(t)
  \simeq
  \mathcal J\!\left(\frac{1}{T_c}-\frac{1}{T_b}\right)t,
\label{eq:prePage}
\end{equation}
yielding a linear pre-Page proxy whose slope is set entirely by the two-horizon
thermodynamics. (Greybody effects or nonconformal corrections modify only the
numerical coefficients.)

\subsubsection{Refined Page condition}

The Page time $t_{\rm Page}$ is defined as the moment when the generalized entropies of
the two competing saddles become equal,
\[
S_{\rm gen}^{(\text{no--island})}(t_{\rm Page})
= S_{\rm gen}^{(\text{island})}(t_{\rm Page})\,.
\]
At the semiclassical level this equality is well approximated by two standard
assumptions:
(i) the no--island entropy growth is estimated by the coarse--grained proxy
$S_{\rm proxy}$, and
(ii) the island saddle is dominated by the black hole’s
Bekenstein--Hawking entropy $S_b$.
Under these approximations, the Page condition simplifies to
\begin{equation}
S_{\rm proxy}(t_{\rm Page}) \simeq S_b(t_{\rm Page})\,.
\label{eq:PageEquality}
\end{equation}

Using $\dot M=-\mathcal J(M)$ and
$\dot S_{\rm proxy}=\mathcal J(M)\!\left(\tfrac{1}{T_c(M)}-\tfrac{1}{T_b(M)}\right)$,
one can eliminate $\mathcal J(M)$ to obtain
\begin{equation}
\frac{dS_{\rm proxy}}{dM}
  = -\!\left(\frac{1}{T_c(M)}-\frac{1}{T_b(M)}\right),
\end{equation}
which integrates to
\begin{equation}
S_{\rm proxy}(M)
 = \int_{M}^{M_0}\!\!\left(\frac{1}{T_c(m)}-\frac{1}{T_b(m)}\right)dm.
\label{eq:SradOfM}
\end{equation}
The Page condition \eqref{eq:PageEquality} therefore becomes
\begin{equation}
\int_{M_{\rm Page}}^{M_0}\!\!\Big(\tfrac{1}{T_c}-\tfrac{1}{T_b}\Big)dm
 \;=\; S_b(M_{\rm Page}),
\label{eq:PageCondition}
\end{equation}
with $S_b(M)=\pi r_b(M)^2/G_4$ and $r_b(M)$ defined by $\xi(r_b;M,\Lambda)=0$.

The corresponding Page time follows from the adiabatic relation
\begin{equation}
t_{\rm Page}
 = \int_{M_{\rm Page}}^{M_0}\!\frac{dm}{\mathcal J(m)}
 = \frac{48\pi}{N}\!
 \int_{M_{\rm Page}}^{M_0}\!\frac{dm}{\kappa_b^2(m)-\kappa_c^2(m)} \, .
\label{eq:tPageIntegral}
\end{equation}
Because $T_b$ increases as the black hole shrinks,
$S_{\rm proxy}$ grows faster than the frozen-temperature estimate predicts,
so $S_b(t_{\rm Page})<S_b(0)$ and the true Page time occurs earlier
than the naive constant-slope approximation.

\begin{figure}[t]
  \centering
  \includegraphics[width=14cm]{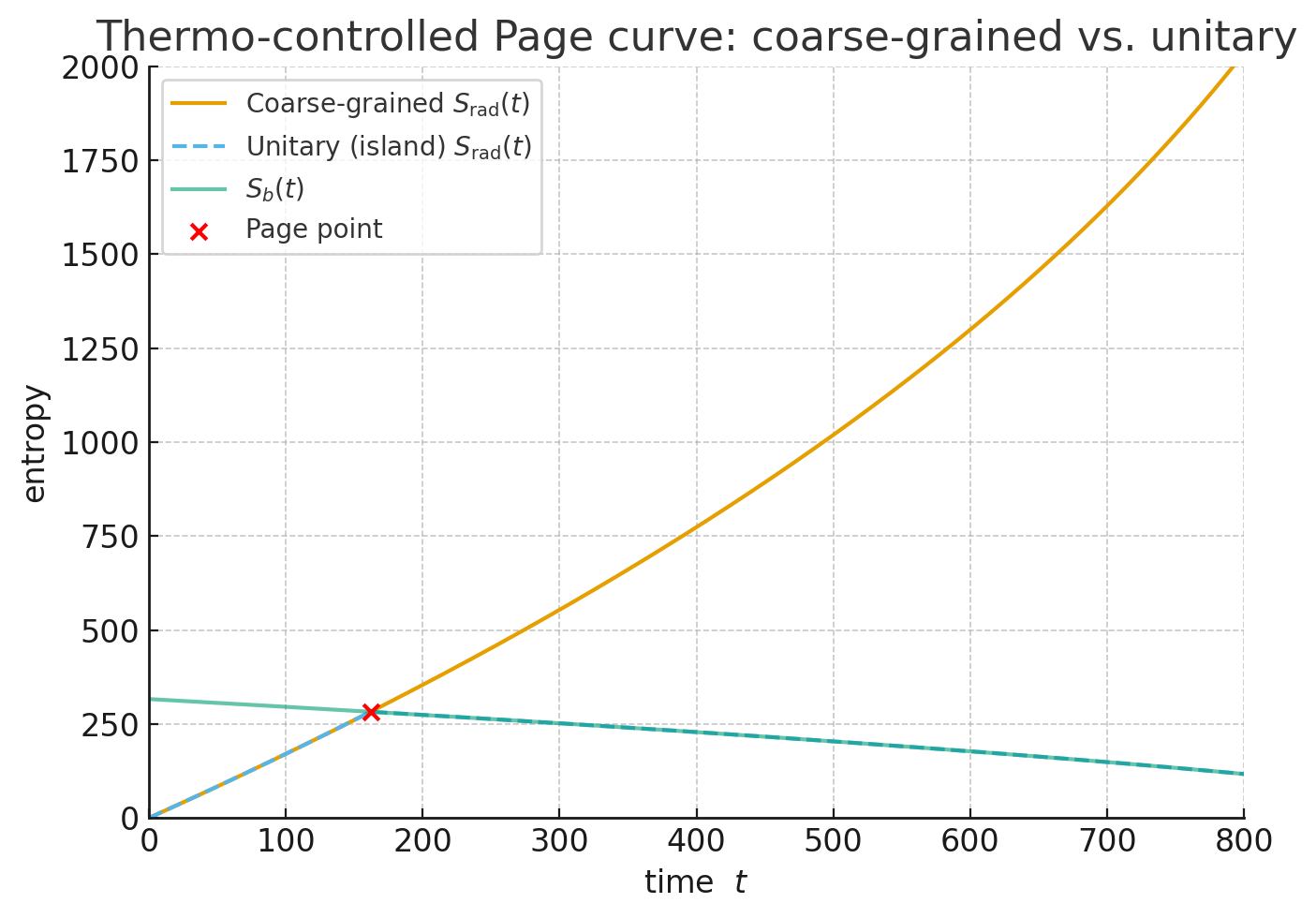}
  \caption{\justifying Thermo-controlled Page-curve proxy for Schwarzschild--de~Sitter.
  The solid (yellow) curve shows the coarse-grained entropy-production proxy $S_{\rm proxy}(t)$ obtained from the steady-state flux [Eqs.~\eqref{eq:J-Ts}, \eqref{eq:prePage}]. 
  The dashed (blue) curve depicts the heuristic island-bounded estimate $S_{\rm proxy}^{\rm(island)}(t)=\min\{S_{\rm proxy}(t),S_b(t)\}$, which turns over at the Page point (red marker) where the island saddle dominates [Eqs.~\eqref{eq:PageEquality}–\eqref{eq:tPageIntegral}]. 
  The (green) curve is the black-hole entropy $S_b(t)$. 
  The cosmological horizon enters through $T_c$ and the anomaly-induced flux; no additional fit parameters are introduced. 
  (Parameters shown are representative; the qualitative behavior is robust for all $0<9M^2\Lambda<1$.)}
  \label{fig:PageCurve}
\end{figure}

Figure~\ref{fig:PageCurve} contrasts the coarse-grained thermodynamic proxy,
obtained by integrating the adiabatic relation
\[
\dot S_{\rm proxy}(t)
 = \mathcal{J}(M(t))
   \!\left(\frac{1}{T_c(M(t))}-\frac{1}{T_b(M(t))}\right),
\qquad
\dot M(t)=-\mathcal{J}(M(t)),
\]
with $\mathcal{J}(M)$ given by Eq.~\eqref{eq:J-Ts},
against the heuristic island-bounded estimate
\[
S_{\rm proxy}^{\rm(island)}(t)=\min\{S_{\rm proxy}(t),S_b(t)\}.
\]
The turnover at $t_{\rm Page}$ (red marker) realizes the saddle switch implied by
Eqs.~\eqref{eq:PageEquality}–\eqref{eq:tPageIntegral} in this thermodynamic
proxy estimate.
In this construction, the cosmological horizon is accounted for through $T_c$
and the anomaly-induced flux $\mathcal{J}$, so no additional radiation reservoir
needs to be introduced.
The qualitative features—linear pre-Page growth determined by the two-horizon
temperature contrast, followed by an island-controlled decrease tracking $S_b(t)$—
are insensitive to greybody effects and persist throughout the neutral
Schwarzschild--de~Sitter domain $0<9M^2\Lambda<1$.%
\footnote{The Page point in Fig.~\ref{fig:PageCurve} does not occur when the black hole
has lost half of its initial entropy. In Schwarzschild--de~Sitter spacetime,
the presence of the cosmological horizon and the steady flux modify the entropy balance,
so the proxy condition $S_{\rm proxy}\simeq S_b$ replaces the heuristic
$S_b\simeq \tfrac12 S_b(0)$ of single-horizon models.}

\medskip
Since both $\mathcal J(m)$ and $\big(1/T_c-1/T_b\big)(m)$ increase as $M$ decreases,
freezing them at their initial values underestimates the slope of $S_{\rm proxy}(t)$.
Consequently,
\begin{equation}
  t_{\rm Page}
  \;\le\;
  \frac{S_b(0)}{\ \mathcal J_0\!\left(\frac{1}{T_{c0}}-\frac{1}{T_{b0}}\right)}\,,
\end{equation}
which serves as a conservative upper bound rather than a precise prediction.

A few comments are in order:
(i) The slope in \eqref{eq:prePage} matches the $1{+}1$-dimensional CFT
nonequilibrium steady-state relation
$\mathcal{J}=\tfrac{\pi c}{12}(T_b^2-T_c^2)$ with central charge $c=N$, now obtained directly from the Polyakov stress tensor in a curved two-horizon geometry.
This provides the first fully analytic, anomaly-induced demonstration that the generalized second law is satisfied in Schwarzschild–de Sitter evaporation.
(ii) Equations \eqref{eq:PageCondition}–\eqref{eq:tPageIntegral} show that
increasing $N$ or the temperature contrast $(T_b-T_c)$ shortens
$t_{\rm Page}$, while larger initial entropy $S_b(0)$ delays it.
(iii) A microscopic QES derivation would extremize
$S_{\rm gen}[\partial\mathcal I]
 =\tfrac{X(\partial\mathcal I)}{4G_2}+S_{\rm out}(\mathcal R\cup\mathcal I)$;
the present estimate does not replace that calculation, but uses the
thermodynamic proxy to locate the expected turnover scale.

The resulting curve should be read as a Page-like thermodynamic estimate within
the semiclassical framework. A genuine microscopic Page curve would require the
full fine-grained entropy functional and island saddle analysis, which lies
beyond the present proxy calculation.

Although a full evaluation of $S_{\rm out}$ and the exact island geometry
is beyond the present scope, our solvable SdS model
provides a controlled solvable setting for exploring Page-like behavior
in cosmological spacetimes, extending the modern island paradigm to
multi-horizon backgrounds~\cite{Geng:2021wcq}.

\section{Conclusions}
Here we summarize our main conclusions and discuss possible future research directions. While our results accord with general expectations from black-hole thermodynamics, the framework developed here provides a fully analytic, backreacted realization of Schwarzschild–de~Sitter evaporation, demonstrating explicitly that the anomaly-induced flux, horizon Clausius relations, and generalized second law coexist consistently within a single solvable model.

\subsection{Main conclusions}
\begin{itemize}
\item The conserved Hawking flux $\mathcal{J}$ (Sec.~\ref{sec:framework})
drives the semiclassical evolution of the mass according to
Eq.~\eqref{eq:massloss}, representing steady energy transport from the
black-hole to the cosmological horizon.  Since $\Delta>0$ for all
neutral SdS configurations, the flux is always outward and the
black-hole mass decreases monotonically.

\item The system generically occupies a steady
nonequilibrium state with $\kappa_b\!>\!\kappa_c$, in which radiation
flows from the hotter to the cooler horizon.  True thermal equilibrium
occurs only when $\kappa_b=\kappa_c$, i.e.\ at the Nariai boundary
where both surface gravities vanish and $\mathcal{J}=0$.

\item The $(M,\Lambda)$ phase diagram shows that the entire static-patch
region $0<9M^2\Lambda<1$ lies in the evaporation regime.
The Nariai configuration marks the unique zero-flux fixed point where
the horizons coincide.

\item Locally measured quantities, such as temperature, flux, and energy
density, obey Tolman redshift and behave as expected for a finite
thermal cavity bounded by walls at unequal temperatures.

\item The GSL holds throughout the semiclassical evolution:
although $S_b$ decreases, $S_c$ increases more rapidly, ensuring that
the total generalized entropy $S_{\rm gen}=S_b+S_c$ grows
monotonically.

\item Extending beyond thermodynamics, the same framework gives a
Page-like thermodynamic proxy: the coarse-grained entropy-production
estimate grows linearly at a rate fixed by the temperature contrast and
steady flux, and the expected island turnover scale can be estimated by
comparing this proxy with \(S_b(t)\). A microscopic Page curve would require
a full QES extremization.
\end{itemize}

These results highlight the effectiveness of the
two-dimensional anomaly-induced framework for capturing both the
thermodynamic and informational aspects of black holes
with multiple horizons. 
Within the semiclassical approximation the mass decreases monotonically and the
black hole evaporates completely, leaving an empty de~Sitter static patch.
Once $M$ becomes comparable to the Planck mass, however, the effective
two-dimensional description ceases to be reliable, and phenomena such as a
final burst of Hawking radiation or a short-lived Planck-scale remnant would
require a full quantum-gravity treatment beyond the present analysis.

In contrast to fully four-dimensional
analyses, which are often numerical or state-dependent, the present model provides
analytic control of flux, backreaction, entropy flow, and Page-like
behavior while remaining conceptually transparent.  The effective
action encodes evaporation, entropy production, and information
transfer within a solvable setting that isolates the dominant $s$-wave
sector.  The inclusion of greybody corrections or non-conformal matter, while beyond the exact 2D anomaly framework used here, would provide natural directions for extending the model toward more realistic four-dimensional physics.
\medskip
\subsection{Future directions}
\begin{itemize}
\item \emph{Charged and nonsingular black holes in de~Sitter.}
The charged Reissner--Nordstr\"om--de~Sitter extension of the present
neutral analysis is developed in a sequel paper~\cite{Easson:2026kng}.
There the anomaly-induced heat flux is supplemented by electromagnetic work
and Schwinger discharge, leading to coupled mass--charge evolution. Extending
the same framework to regular geometries, such as Bardeen or Hayward black
holes in de~Sitter space, would enable a unified treatment of inner and
cosmological horizons within the anomaly-induced semiclassical theory.

\item \emph{Entanglement islands and quantum extremal surfaces.}
A full QES analysis using the explicit Polyakov field and stress tensor
developed here could determine the precise island location and Page
time, extending recent island-paradigm results to cosmological
spacetimes with multiple horizons.

\item \emph{Beyond the conformal anomaly.}
Including massive or non-conformal fields, through perturbative or
numerical treatments, would test the universality of the
anomaly-driven results and connect with realistic matter sectors.

\item \emph{Cosmological horizon thermodynamics.}
The SdS patch behaves as a finite static universe bounded by two
thermal walls.  Investigating the thermodynamic properties of the
cosmological horizon from the interior, including its heat capacity and
response to backreaction, may offer insight relevant to inflationary
and late-time cosmology.
\end{itemize}

In summary, the anomaly-corrected SdS model offers
a uniquely controlled setting in which quantum backreaction,
nonequilibrium thermodynamics, and information recovery can all be
treated analytically.  It establishes a foundation for exploring
quantum steady states and Page-like behavior in spacetimes with
multiple horizons, and provides a benchmark for future extensions to
charged, rotating, regular or fully quantum-gravitational systems.

\acknowledgments
It is a pleasure to thank Lars Aalsma, Paul Davies, Cynthia Keeler, Phillip Levin, Don Page, Marija Tomasevic and Tanmay Vachaspati for useful discussions. This work is supported by the U.S. Department of Energy, Office of High Energy Physics, under Award Number DE-SC0019470.
\\

\newpage
\appendix
\section{Exact quadrature reduction and the static momentum constraint}
\label{app:quadratures}

Here we collect only the pieces of the semiclassical system needed for the
static momentum constraint and first-integral interpretation used in the main
text. The gravitational sector is the SRG action \eqref{eq:act2}; the Polyakov
term may be localized by introducing an auxiliary field $\psi$ satisfying
$\Box\psi=R$. In a static geometry of the form
\begin{equation}
ds^2 = -f(r)\,dt^2 + f(r)^{-1}dr^2,
\label{app:staticmetric}
\end{equation}
with \(X=X(r)\), the off--diagonal
semiclassical Einstein equation enforces a vanishing total momentum density,
\(T_{rt}=0\). Thus an \emph{exact} flux--carrying solution cannot be strictly
static; the nonzero Unruh--dS flux is realized in the quasi--static regime used
in the main text (or in a time-dependent ansatz, e.g.\ Eddington--Finkelstein form),
while exact static solutions correspond to zero net flux.

\subsection{Preliminaries in static gauge}

In the static gauge one has \(\sqrt{-g}=1\) and
\begin{equation}
R=-\,f''(r),\qquad
\Box \Phi = \partial_r\!\big(f\,\Phi'(r)\big)\quad\text{for any scalar }\Phi(r).
\label{app:Rbox}
\end{equation}
\noindent\textit{Notation.} A prime on \(f\), \(X\), or \(\psi\) denotes
\(\partial_r\), except when explicitly applied to a function of \(X\).

\subsection{First integrals}

The localized Polyakov equation $\Box\psi=R$ gives
\begin{equation}
(f\,\psi')' \;=\; R \;=\; -\,f''(r)
\quad\Rightarrow\quad
f\,\psi' \;=\; -\,f' + C_\psi,
\label{app:psifirst}
\end{equation}
The integration constant \(C_\psi\), together with the state functions
\(t_\pm\), is fixed by requiring the renormalized stress tensor to be regular
in the appropriate future Kruskal frames. Its explicit value is not needed for
the flux-balance argument.

The remaining static SRG equations can be organized by the usual Casimir
first integral. In the absence of net flux the static solution is specified by
the Casimir mass, the state data, and the regularity condition on $\psi$.
This first-integral structure is the only input from the static problem needed
below; the flux-carrying evolution is treated instead in EF gauge in
Appendix~\ref{app:EF}.

\paragraph*{State choice and flux---}Stationarity would make
$T^{r}{}_{t}$ radially constant in the static gauge.  The momentum constraint
below fixes this constant to zero for any strictly static solution.  The
nonzero Unruh--de~Sitter flux is instead represented by the conserved SRG flux
$\mathcal J$ in the quasistationary EF treatment of Appendix~\ref{app:EF}.

\subsection{Static momentum constraint and a no-go for exact static flux}
\label{app:nogo}

In a static metric ansatz with $g_{rt}=0$ and $X=X(r)$, the mixed
gravitational equation has no source from the SRG geometric sector: the
off--diagonal metric component and the mixed derivative of $X$ both vanish.
The momentum constraint therefore enforces
\begin{equation}
T^{(\psi)}_{rt} + T^{\rm(state)}_{rt}=0,
\qquad
T^{r}{}_{t}=0.
\label{app:mixzero}
\end{equation}
It is convenient to define $J:=-T^{r}{}_{t}$ as the flux in the static limit.
The momentum constraint~\eqref{app:mixzero} thus requires $J=0$, showing that
no exact static solution of the semiclassical SRG equations can support a
nonzero energy flux.
Although stationarity implies $\nabla_\mu T^{\mu}{}_{t}=0$ (so $T^{r}{}_{t}$ is
radially constant), the momentum constraint fixes that constant to zero in any
strictly static geometry.
A nonvanishing, radially constant flux $\mathcal{J}\neq0$ becomes consistent only when
strict staticity is relaxed---for instance, in an adiabatic
Eddington--Finkelstein gauge where the mass parameter evolves slowly, $M(v)$, with
$\dot M=-\mathcal{J}(M,\Lambda)$, or more generally in a stationary (nonstatic) ansatz
containing $g_{vr}$ or $\partial_v$ terms that balance the $(v,r)$ field equation.
This quasi--stationary setting is the framework adopted in the main text.

\subsection{Quadratures and horizon regularity (static, zero-flux case)}
\label{app:quads}

In the strictly static case \(\mathcal{J}=0\), regularity on \emph{both}
future horizons corresponds to a Hartle--Hawking--dS--type configuration,
which for neutral SdS exists only at the Nariai limit. By contrast, the
Unruh--dS state (regular on the future horizons but carrying nonzero flux)
requires a time-dependent ansatz such as Eddington--Finkelstein form or,
equivalently, a quasi-static treatment in which \(\dot M = -\mathcal{J}(M,\Lambda)\)
at leading adiabatic order.

The static first-integral problem may be reduced to quadratures once a
monotone branch \(X(r)\) (or equivalently \(r(X)\)) is chosen and the Casimir
mass is fixed. Regularity of the renormalized stress tensor on the future
horizons \(\mathcal{H}_b^+\) and \(\mathcal{H}_c^+\) fixes the state constants
\(t_u\) and \(t_v\), together with \(C_\psi\), for any zero-flux static state.
In the strictly static case,
\eqref{app:mixzero} enforces \(\mathcal{J}=0\), i.e.\ no net flux/equilibrium.
For neutral SdS a global thermal equilibrium occurs only at the Nariai limit:
in the original static normalization $\kappa_b=\kappa_c=0$, while the
near-horizon Nariai geometry uses a rescaled time with its own finite
temperature.

\medskip
\noindent\textit{Summary}---Static solutions necessarily have $\mathcal{J}=0$;
the steady Unruh--de~Sitter flux arises only when strict staticity is relaxed,
as in the quasi--stationary Eddington--Finkelstein formulation used in the main text.

\subsection{Dimensional-reduction anomaly}

The present analysis employs the standard spherically reduced
two--dimensional dilaton--gravity action augmented by the Polyakov term,
which correctly captures the structure of the trace anomaly of \(N\) conformal matter fields but omits the
so--called \emph{dimensional-reduction anomaly}: additional state-dependent terms that arise when
integrating out angular modes in the four-dimensional theory.
These corrections can modify the quantitative relation between the
4D and 2D fluxes, rescaling the magnitude of \(\mathcal{J}\) or shifting
the effective dilaton potential. However, they do \emph{not} affect the sign structure of the flux or
the inequalities central to our conclusions. The positivity of \(\Delta =\kappa_b^2-\kappa_c^2\), the direction of
energy flow \(\mathcal{J}\propto\Delta(M)\), and the non-negativity of the
generalized entropy production rate
\(\dot S_{\mathrm{gen}}\sim\mathcal{J}(1/T_c-1/T_b)\) follow purely from
geometric relations among the Schwarzschild--de~Sitter horizons and thus
remain unchanged under any overall normalization of the stress tensor.
In this sense, the dimensional-reduction anomaly alters \emph{how fast}
evaporation proceeds but not \emph{which way} it goes or whether the
generalized second law holds. This line of thinking is further justified by results from the four dimensional case \cite{Hollands:2019whz}.

\section{Adiabatic evolution in Eddington--Finkelstein gauge and the mass--loss law}
\label{app:EF}
This appendix complements Appendix~\ref{app:quadratures} by demonstrating that a nonzero
constant flux $\mathcal{J}$ is compatible with the semiclassical equations at leading
order in a slow-time (adiabatic) expansion. In advanced Eddington--Finkelstein (EF)
coordinates, time dependence enters the mixed $(v,r)$ components of the equations of
motion, and the mass-loss relation
\begin{equation}
\frac{dM}{dv} \;=\; -\,\mathcal{J}
\end{equation}
follows directly from a covariant Casimir (mass function) balance law. We assume
$\partial_v f \ll \kappa f$ and keep $\mathcal{O}(N)$ backreaction from the Polyakov
anomaly.

\subsection{EF ansatz and basic identities}

To describe flux solutions it is convenient to work in advanced
Eddington--Finkelstein (EF) coordinates $(v,r)$ with
\begin{align}
ds^2 &= -\,f(v,r)\,dv^2 + 2\,dv\,dr, \nonumber \\
X &= X(r) \quad (\text{for the Schwarzschild--de~Sitter reduction, } X=r^2), \nonumber \\
\psi &= \psi(v,r),
\label{eq:EFmetric}
\end{align}
where the classical metric function is
\[
f_{\rm cl}(M,\Lambda;r)
   = 1 - \frac{2M}{r} - \frac{\Lambda}{3}\,r^2,
\]
and we allow the mass parameter $M=M(v)$ to vary slowly,
incorporating $\mathcal{O}(N)$ backreaction corrections in~$f$.

\medskip
\noindent
\textit{Comment on coordinates---}In the EF chart the metric is non-diagonal, so its inverse has both
$g^{rv}=1$ and $g^{rr}=f(v,r)$.
These components are distinct from those appearing in the diagonal
(static) gauge of Eq.~\eqref{app:staticmetric};
no mixing of coordinate systems occurs.
The same function $f$ appears because it represents the
redshift factor of the geometry in either chart.

\medskip
\noindent
Useful identities in this gauge are
\begin{align}
&\sqrt{-g}=1,\qquad
  g^{vv}=0,\qquad
  g^{vr}=g^{rv}=1,\qquad
  g^{rr}=f, \\[4pt]
&\Box \Phi
   = 2\,\partial_v\partial_r \Phi
     + \partial_r\!\big(f\,\partial_r \Phi\big),
\label{eq:boxEF}\\[4pt]
&R
   = -\,\partial_r^2 f(v,r).
\label{eq:RicciEF}
\end{align}

The localized Polyakov equation $\Box\psi=R$ therefore becomes
\begin{equation}
2\,\partial_v\psi' + (f\psi')'
   \;=\; -\,f''(v,r).
\label{eq:EFpsiEOM}
\end{equation}
Here primes denote $\partial_r$ and we use dots for $\partial_v$.

\subsection{Slow-time equations and state flux}
The mixed equations in EF gauge contain the slow-time terms that make a
nonzero flux compatible with the semiclassical equations.  For the present
purpose the most economical route is to use the covariant SRG Casimir balance
law, which directly relates the change in the mass function to the conserved
Killing flux.  This avoids introducing a separate pure-$\tilde V$
gravitational system and keeps the derivation in the same SRG conventions as
the main text.

\subsection{Mass function (Casimir) balance and \texorpdfstring{$\dot M = -\mathcal J$}{Mdot = -J}}

We now derive the mass-loss law from the covariant balance law for the
areal-radius SRG mass function. Define $Q(X)$ by
$Q'(X)=-\,U(X)$ and $I(X):=e^{-Q(X)}$. With the conventions of Sec.~II, the
mass function (or dilaton Casimir) is
\begin{equation}
\mathcal C \;=\; w(X)-\frac{1}{4}e^{Q(X)}(\nabla X)^2,
\qquad w'(X) \;=\; \frac{1}{2}\,\tilde V(X).
\label{eq:Cmass}
\end{equation}
For \(X=r^2\), \(e^Q=1/r\), and \((\nabla X)^2=4r^2 f\), this gives
\(\mathcal C=w-rf=2G_4M\).
In the presence of matter, the exact balance law reads
\begin{equation}
\nabla_\mu \mathcal C \;=\; 2G_4\,I(X)\,T_{\mu\nu}\,\nabla^\nu X.
\label{eq:Cbalance}
\end{equation}
With $X=X(r)$ in the EF metric \eqref{eq:EFmetric}, one has $\nabla^\nu X=g^{\nu r}X'$ and
$T_{v\nu}\nabla^\nu X = X'\,T^{r}{}_{v}$, so
\begin{equation}
\partial_v \mathcal C \;=\; 2G_4\,I(X)\,X'(r)\,T^{r}{}_{v}.
\end{equation}
Identifying $\mathcal C=2G_4 M(v)$ (up to an additive constant) yields the general mass–loss formula
\begin{equation}
 \dot M(v) \;=\; I(X)\,X'(r)\,T^{r}{}_{v}\ .
\label{eq:mdot-general}
\end{equation}
In the original SRG frame ($U=1/(2X)\Rightarrow Q'=-1/(2X)\!,\ I=e^{-Q}=\sqrt{X}=r$) one has
$I(X)X'(r)=r\cdot 2r=2r^2$, hence
\begin{equation}
\dot M(v) \;=\; 2\,r^2\,T^{r}{}_{v}.
\label{eq:mdot-srg}
\end{equation}
It is convenient to define the EF flux (which is radially conserved at leading adiabatic order)
\begin{align}
\mathcal J := -\,I(X)\,X'(r)\,T^{r}{}_{v}, \label{eq:Deltadef1} \\[4pt]
\dot M(v) = -\,\mathcal J. \label{eq:Deltadef2}
\end{align}
For \(X=r^2\) and \(I=r\), this convention gives
\begin{equation}
T^{r}{}_{v}=-\frac{\mathcal J}{2r^2},
\end{equation}
so positive \(\mathcal J\) denotes outward Killing-energy flux.

In Eddington--Finkelstein gauge, the semiclassical equations therefore admit an
adiabatically evolving solution in which the constant Unruh--de~Sitter flux
$\mathcal J$ drives the mass according to Eq.~\eqref{eq:Deltadef2}.
This derivation explicitly shows how the time-dependent ansatz circumvents
the static momentum constraint of Appendix~\ref{app:quadratures} and provides
a covariant foundation for the mass--loss law used in the main text.

\section{Nariai limit in the two--dimensional reduction}
\label{app:Nariai}

In the spherical reduction of four-dimensional Einstein gravity,
\begin{equation}
ds^2_{(4)} = g_{ab}(x)\,dx^a dx^b + r^2(x)\, d\Omega_2^2,
\end{equation}
the two-dimensional dilaton field is $X=r^2$.  The Schwarzschild--de~Sitter
(SdS) spacetime possesses two horizons, the black-hole and cosmological
horizons $(r_b,r_c)$, defined by the zeros of
\begin{equation}
\xi(r;M,\Lambda)=1-\frac{2M}{r}-\frac{\Lambda r^2}{3}.
\end{equation}
The \emph{Nariai limit} corresponds to the extremal configuration in which these
horizons coincide,
\begin{equation}
r_b=r_c=r_N=\frac{1}{\sqrt{\Lambda}},\qquad
M=M_N=\frac{1}{3\sqrt{\Lambda}},\qquad
X_N=r_N^2=\frac{1}{\Lambda}.
\end{equation}
At $r=r_N$ one has 
\begin{equation}
\xi(r_N)=0=\xi'(r_N),
\end{equation}
so that the surface gravities vanish, $\kappa_b=\kappa_c=0$, and the
geometry factorizes as $\mathrm{dS}_2\times S^2$.

\subsection{Near-horizon scaling and coordinates}

To obtain the Nariai metric explicitly, we expand the four-dimensional
SdS metric near the degenerate horizon and introduce rescaled coordinates
that remain finite as the double root forms.  Let $\varepsilon$ parameterize
the deviation from extremality such that $r_c-r_b\propto\varepsilon$.
Define dimensionless near-horizon coordinates
\begin{equation}
r = r_N\big(1+\varepsilon\, y\big),
\qquad
t = \frac{\tau}{\varepsilon},
\label{eq:NariaiScaling}
\end{equation}
and take the limit $\varepsilon\to0$ with $(\tau,y)$ held fixed.
The variable $y\in(-1,1)$ is a dimensionless static coordinate on the
two-dimensional de~Sitter factor, with Killing horizons at $y=\pm1$,
while $\tau$ is the corresponding static time.
The $1/\varepsilon$ rescaling ensures a nontrivial finite limit since
the surface gravity vanishes linearly with $\varepsilon$.
Although $t=\tau/\varepsilon$ formally diverges as $\varepsilon\to0$, this simply
reflects the infinite redshift at the degenerate horizon; the rescaled time
$\tau$ remains finite and defines the proper static time coordinate of the
Nariai patch.

\subsection{Two-dimensional Nariai geometry}

Substituting \eqref{eq:NariaiScaling} into the reduced two-dimensional metric
$ds^2_{(2)}=-\xi(r)\,dt^2+\xi(r)^{-1}dr^2$ (with $\xi\equiv f$)
and expanding to leading order in
$\varepsilon$ yields
\begin{equation}
ds^2_{(2)}=\frac{1}{\Lambda}
\!\left[-(1-y^2)\,d\tau^2 + \frac{dy^2}{1-y^2}\right],
\label{eq:Nariai2D}
\end{equation}
which is the static patch of two-dimensional de~Sitter space
$\mathrm{dS}_2$ with curvature radius $\ell_2=1/\sqrt{\Lambda}$.
The corresponding Ricci scalar is
\begin{equation}
R^{(2)} = 2\Lambda,
\end{equation}
and the Killing horizons of the metric lie at $y=\pm1$.

At the same time, the dilaton approaches a constant,
\begin{equation}
X \;=\; r^2 \;\longrightarrow\; X_N \;=\; \frac{1}{\Lambda},
\end{equation}
so the internal $S^2$ factor freezes at radius $r_N$.
The full four-dimensional Nariai geometry is therefore
\begin{equation}
\mathrm{dS}_2(\ell_2^2=1/\Lambda)\,\times\,S^2(r_N^2=1/\Lambda).
\end{equation}

\subsection{Interpretation in the 2D dilaton theory}

In the two-dimensional dilaton-gravity framework, the Nariai configuration
corresponds to a \emph{double root} of the Killing function $\xi(X)$,
\begin{equation}
\xi(X_N)=0=\xi'(X_N)
\quad\Longleftrightarrow\quad
w'(X_N)=0.
\end{equation}
At this point, the static patch reaches its degenerate boundary: the
black-hole and cosmological horizons coincide, and the surface gravities
defined with respect to the original SdS static Killing field vanish.
After the near-horizon rescaling, however, the limiting two-dimensional
metric is $\mathrm{dS}_2$ with constant dilaton and has its own finite
temperature with respect to the rescaled Nariai time.  This configuration
represents the limiting equilibrium state of neutral
Schwarzschild--de~Sitter spacetime in both the four-dimensional and
two-dimensional descriptions.

\section{Curvature scalar and horizon regularity}
\label{app:curvature}

Here we record the curvature regularity criterion used in the quasistationary
analysis. This appendix does not independently prove existence or smoothness of
the full backreacted solution; rather, it states the geometric condition under
which the two-dimensional Ricci scalar remains finite at the horizons.

\subsection{Ricci scalar in static (or EF) gauge}

In the static gauge
\begin{equation}
ds^2 = -f(r)\,dt^2 + f(r)^{-1}dr^2,
\end{equation}
the two-dimensional Ricci scalar is
\begin{equation}
R(r) = -\,f''(r).
\label{eq:Rfromf_app}
\end{equation}
(Equivalently, in the corresponding ingoing Eddington--Finkelstein form
$ds^2=-\xi(r)\,dv^2+2\,dv\,dr$, one has $R=-\,\xi''(r)$, since
$f(r)\equiv\xi(r)$ for the static solution.)
Thus, if the quasistationary semiclassical metric function $f(r)$ is \(C^2\)
through the horizons, the two-dimensional Ricci scalar is finite there.

\subsection{Finiteness at the horizons}

Let $r=r_h$ be a (nondegenerate) Killing horizon, $f(r_h)=0$ with
$f'(r_h)=2\kappa_h\neq0$. Expanding near $r_h$,
\begin{equation}
f(r) = 2\kappa_h (r-r_h) + \tfrac12 f''(r_h)(r-r_h)^2
        + \mathcal{O}\!\big((r-r_h)^3\big),
\end{equation}
the curvature is
\begin{equation}
R(r_h) = -\,f''(r_h),
\end{equation}
which is finite provided the near-horizon expansion of $f(r)$ is \(C^2\). This
is the regularity criterion imposed on the quasistationary semiclassical
metric.

\medskip
\noindent\emph{Degenerate limit---}In the Nariai limit $\kappa_h\!\to\!0$ the linear term
vanishes, but the analysis proceeds with the next term in the expansion;
$R=-f''(r)$ remains finite.

\subsection{Classical limit and backreaction corrections}

Since the anomaly-induced backreaction enters linearly in the Polyakov coefficient,
we may write
\begin{equation}
R(r)=R_{\rm cl}(r)+\delta R(r),\qquad \delta R(r)=\mathcal O(N)\,,
\end{equation}
with $R_{\rm cl}(r)=-\,f_{\rm cl}''(r)$ for classical Schwarzschild--de~Sitter.
If the regular Unruh--de~Sitter state is represented by a \(C^2\)
quasistationary metric function on $[r_b,r_c]$, then $\delta R(r)$ is finite and
$R(r)\to R_{\rm cl}(r)$ smoothly as $N\to0$. Thus the two-dimensional Ricci
scalar is regular throughout the static patch, including both horizons, provided
this metric regularity criterion is satisfied.

\bibliography{2dbhs}

\end{document}